\documentclass[a4paper,11pt]{article}
\pdfoutput=1 
\usepackage{jheppub} 
\usepackage[numbers,sort&compress]{natbib}
\usepackage{amssymb}   
\usepackage{slashed}   
\usepackage{amsmath}   

\usepackage{graphicx}
\usepackage{dcolumn}
\usepackage{bm}

\usepackage{hyperref}
\usepackage{color}
\usepackage{slashed}
\usepackage{multirow}

\usepackage{subfigure}
\usepackage{float}
\usepackage{amsmath}
\usepackage{bbm}

\newcounter{YJC}

\begin{document}

\title{Quantum simulation of the phase transition of the massive Thirring model} 


\author[a]{Jia-Qi Gong}
\author[a,b,1]{Ji-Chong Yang,\note{Corresponding author.}}
\affiliation[a]{Department of Physics, Liaoning Normal University, No. 850 Huanghe Road, Dalian 116029, China}
\affiliation[b]{Center for Theoretical and Experimental High Energy Physics, Liaoning Normal University, No. 850 Huanghe Road, Dalian 116029, China}

\emailAdd{g15124202306@163.com}
\emailAdd{yangjichong@lnnu.edu.cn}

\abstract{
Recent advancements in quantum computing technology have enabled the study of fermionic systems at finite temperature via quantum simulations.
This presents a novel approach to investigating the chiral phase transition in such systems.
Among these, the quantum minimally entangled typical thermal states~(QMETTS) algorithm has recently attracted considerable interest. 
The massive Thirring model, which exhibits a variety of phenomena at low temperatures, includes both a chiral phase transition and a topologically non-trivial ground state. 
It therefore raises the intriguing question of whether its phase transition can be studied using a quantum simulation approach. 
In this study, the chiral phase transition of the massive Thirring model and its dual topological phase transition are studied using the QMETTS algorithm.
Numerical results are obtained on a classical computer simulating circuit-based quantum computations.
The results show that QMETTS is able to accurately reproduce the phase transition and thermodynamic properties of the massive Thirring model.
}

\maketitle


\section{\label{sec:1}Introduction}

In recent years, significant advances have been made in quantum computing technology~\cite{Arute:2019zxq}. 
It is expected that the development of quantum computing will profoundly impact the study of physics in the future.
Although quantum computing remains in the noisy intermediate-scale quantum~(NISQ) era~\cite{Preskill:2018jim,Chen:2021num}, research in high energy physics using quantum algorithms~\cite{Fang:2024ple,Scott:2024txs,Roggero:2018hrn,Roggero:2019myu,Atas:2021ext,Lamm:2019uyc,Li:2021kcs,Perez-Salinas:2020nem,Jordan:2011ci,Mueller:2019qqj,Guan:2020bdl,Wu:2021xsj,Zhang:2023ykh,Zhang:2024ebl,Zhu:2024own,Fadol:2022umw,Wu:2020cye,Terashi:2020wfi,Yang:2024bqw}, particularly quantum simulations~\cite{Feynman:1981tf,Georgescu:2013oza,Bauer:2022hpo,Carena:2022kpg,Gustafson:2022xdt,Lamm:2024jnl,Carena:2024dzu,Li:2023vwx,Cui:2019sfz,Zou:2021pvl,Echevarria:2020wct}, is rapidly advancing.

Quantum simulation has also been proposed to address many-body systems at finite temperature, using methods such as quantum imaginary time evolution~(QITE) algorithms.
One notable QITE-based method is the Quantum Minimally Entangled Typical Thermal States~(QMETTS) algorithm~\cite{Motta:2019yya}, which has recently been applied to the study of the chiral phase transition~\cite{Czajka:2021yll}.
One advantage of quantum simulations is that unlike numerical integration, they use a physical approach. 
As a result, quantum simulations are believed to circumvent the sign problems found in traditional Monte Carlo methods, including the notorious case in lattice quantum chromodynamics~(QCD) at finite chemical potential~\cite{Hasenfratz:1983ba,deForcrand:2009zkb,Gattringer:2016kco}.
The finite chemical potential is important in the study of phase transitions in QCD, because it plays a crucial role in determining the location of the critical endpoint of the phase transition, which is of paramount importance in both experimental and theoretical contexts~\cite{Fukushima:2013rx,Busza:2018rrf,STAR:2010vob,Gupta:2011wh,Andronic:2017pug,Luo:2020pef,STAR:2021iop,Stephans:2006tg,NA49-future:2006qne,Gazdzicki:2008kk,Luo:2017faz,Stephanov:2004wx,Fodor:2004nz,Gavai:2004sd}. 
In addition, there are other sign problems in the study of quark matter that arise under extreme conditions~\cite{Yang:2022zob,Yang:2023vsw,Yang:2023zzx}. 
Therefore, the potential of quantum simulations to effectively study systems at finite temperatures is an avenue that warrants further investigation.

In this work, we use the QMETTS algorithm to study the massive Thirring model~\cite{Thirring:1958in}.
The Thirring model is a toy model describing a self interacting fermion field, which has been extensively studied in previous works~\cite{Sachs:1995dm}, including the study of the mass gap where quantum computing has already played a role~\cite{Mishra:2019xbh}.
One of the reasons for the importance of the Thirring model is that it is exactly soluble in the massless case~\cite{Hagen1967}, while the massive case is soluble by the Bethe ansatz~\cite{Korepin_Bogoliubov_Izergin_1993}.
With the possibility of the exhibition of confinement, the Thirring model provides a chance to use a soluble interacting model to study the confinement~\cite{Alvarez-Estrada:1997vmv,PintoBarros:2018bzz}, which still lacks a complete theoretical description.
The simultaneous breaking of the chiral symmetry in the Thirring model is another attractive phenomenon, as it mirrors key features of QCD phase transitions.
Furthermore, the Thirring model is dual to the sine-Gordon theory~\cite{Coleman:1974bu} which possesses topological structures.
This duality enables the simulation of topological soliton dynamics through the Thirring model~\cite{Banuls:2017evv}.
As will be shown, the massive Thirring model can also be dual to the Gross-Neveu~(GN) model~\cite{Gross:1974jv} with a finite chemical potential, thus this work can also serve as another example of using quantum simulation to study a system with a finite chemical potential.

Given the extensive prior research, our aim is to test whether QMETTS can accurately reproduce known results of the Thirring model.
This not only serves to validate the capabilities of the QMETTS, but also provides a detailed account of its practical applications.
On the one hand, the QMETTS is better suited to the study of systems at high temperatures, whereas the massive Thirring model is also interesting in the low-temperature regime. 
On the other hand, the vacuum of the Thirring model is topologically non-trivial.
It is well established that topologically non-trivial ground states cannot be connected to trivial ones by quantum circuits of finite depth in the thermodynamic limit~\cite{Chen:2010gda}. 
However, in finite-volume systems, the question of whether the barriers between different topological sectors can be crossed by QMETTS is another intriguing topic for exploration. 

The remainder of this paper is organized as follows.
In Section~\ref{sec:2}, a brief introduction to the massive thirring model is presented.
The QMETTS algorithm is discussed in Section~\ref{sec:3}.
Section~\ref{sec:4} presents numerical results on the chiral condensation and topological charge obtained by using the QMETTS algorithm.
In section~\ref{sec:5}, we present discussions on the errors.
Section~\ref{sec:6} is a summary of the conclusions.

\section{\label{sec:2}A brief introduction of the massive Thirring model}

The Thirring model describes self interaction between fermions in $d=1+1$ dimension.
The Hamiltonian for the massive Thirring model is,
\begin{equation}
H=\bar{\psi}\left(i \gamma^{1} \partial_{1}+m\right) \psi-\frac{g^{2}}{4} \bar{\psi} \gamma^{\mu} \psi \bar{\psi} \gamma_{\mu} \psi.
\label{eq.2.1}
\end{equation}
In $d=1+1$,
\begin{equation}
\begin{split}
&\gamma _0=\sigma ^z, \;\; \gamma _1=-i\sigma ^y,
\end{split}
\label{eq.2.2}
\end{equation}
where $\sigma ^i$ are Pauli matrices. 
In order to study the chiral condensation, we use the staggered fermion which preserves the chiral symmetry in the massless limit~\cite{Kogut:1974ag,Kluberg-Stern:1983lmr,Morel:1984di}.
For a staggered fermion, the components of a spinor are distributed within two sites.
Let $h=2n$ denote the even sites, the Dirac fermion field $\psi (h)$ is then written as, 
\begin{equation}
\psi(h)=\frac{1}{\sqrt{a}}\left(\begin{array}{c}\chi (h) \\ \chi(h+1) \end{array}\right),
\label{eq.2.3}
\end{equation}
where $\chi(n)$ is the staggered fermion, $a$ is the lattice spacing between even sites, and the factor $1/\sqrt{a}$ is added to make $\chi(n)$ dimensionless.

In this work, the Minkowski and Euclidean cases of interactions are both considered. 
The Hamiltonian in these cases are denoted by $H_M$ and $H_E$, respectively, and are given by,
\begin{equation}
\begin{split}
&H_{E,M}=a\sum _h\left\{\bar{\psi}(h)\left(i \gamma^{1} \partial_{1}+m\right) \psi(h)
-\frac{g^{2}}{4} \left[\left(\bar{\psi}(h) \gamma^0 \psi(h) \right)^2 \pm \left(\bar{\psi}(h) \gamma^1 \psi(h) \right)^2\right]\right\}.
\end{split}
\label{eq.2.4}
\end{equation}

To discretize the derivative of $\psi$, forward and backward finite differences are used for different spinor components,
\begin{equation}
\begin{split}
&\partial _h \psi(h) = \frac{1}{a\sqrt{a}} \begin{pmatrix} \chi (h+2) -\chi (h) \\ \chi (h+1) - \chi (h-1) \end{pmatrix}.
\end{split}
\label{eq.2.5}
\end{equation}

The staggered fermion is then transformed into the Jordan-Wigner representation~\cite{JW},
\begin{equation}
\chi(n)=\frac{\sigma^x(n)-{\rm i} \sigma^y(n)}{2} \prod_{j=0}^{n-1}\left(-{\rm i} \sigma^z(j)\right),
\label{eq.2.6}
\end{equation}
where $\sigma ^i(n)$ are Pauli matrices sitting on sites, note that different from the $\gamma$ matrices in Eq.~(\ref{eq.2.2}), Pauli matrices in Eq.~(\ref{eq.2.6}) do not act in spinor space.

When the total number of sites $N$ is even, the following equivalences can be verified,
\begin{equation}
\begin{split}
a\sum_{h} \bar{\psi}(h) {\rm i} \gamma_{1} \partial_{x} \psi(h)&=
\frac{1}{2 a} \sum_{i=x,y}\left\{\sum_{n}^{N-2}\sigma^i(n) \sigma^i(n+1)
+(-1)^{\frac{N}{2}}\sigma^i(0) \sigma^i(N-1)\prod_{j=1}^{N-2} \sigma^{z}(j)\right\},\\
a \sum_{h} \bar{\psi}(h) \psi(h)&=\sum_{n=0}(-1)^{n} \frac{1+\sigma^{z}(n)}{2},\\
a\sum_{h}\left(\bar{\psi}(h) \gamma_{0} \psi(h)\right)^{2}
&=\frac{1}{a} \sum_{n}\left(\frac{3}{4}+\sigma^{z}(n)+\frac{\sigma^{z}(n) \sigma^{z}(n+1)}{4}\right),\\
a \sum_{h}\left(\bar{\psi}(h) \gamma_{1} \psi(h)\right)^{2}&=-\frac{1}{4 a} \sum_{n}\left(1-\sigma^{z}(n) \sigma^{z}(n+1)\right),\\
a \sum_{h} \bar{\psi}(h) \gamma_{0} \psi(h)&=\sum_{n=0} \frac{1+\sigma^{z}(n)}{2},\\
\end{split}
\label{eq.2.11}
\end{equation}
where the terms with higher orders of $a$ are neglected to keep the translational invariance of the Hamiltonian. 
The omitted terms are shown in Appendix~\ref{sec:ap1}. 
Using,
\begin{equation}
\begin{split}
&a\sum _h (\bar{\psi}(h) \psi(h) )^2=\frac{1}{4a}\sum _n \left(1+\sigma^z(n)\sigma^z(n+1)\right),\\
\end{split}
\label{eq.2.12}
\end{equation}
the massive Thirring model in $d=1+1$ can be related to the GN model with a chemical potential, whose Hamiltonian is,
\begin{equation}
\begin{split}
&H_{GN}(m,g,\mu)=\bar{\psi}\left(i \gamma_{1} \partial_{1}+m\right) \psi
-g(\bar{\psi} \psi)^{2} -\mu \bar{\psi} \gamma_{0} \psi,
\end{split}
\label{eq.2.13}
\end{equation}
where $\mu$ is chemical potential.
Neglecting the constants and $\mathcal{O}(a)$ terms, it can be seen that,
\begin{equation}
\begin{split}
&H_M(m,g)=H_{GN}(m,0,\frac{ag^2}{2}),\\
&H_E(m,g)=H_{GN}(m,\frac{g^2}{2},\frac{ag^2}{2}).\\
\end{split}
\label{eq.2.14}
\end{equation}

It has been shown in Ref.~\cite{Czajka:2021yll} by using QMETTS that there is a chiral phase transition where the chiral symmetry is spontaneously broken.
Therefore, it can be expected that the massive Thirring model will also present a chiral phase transition.
Furthermore, since the interaction effectively acts as a chemical potential, it induces spontaneous symmetry breaking in the massive Thirring model even without an explicit chemical potential.
To investigate such a phenomenon, the chiral condensation as a function of $g$ and temperature $T$ is considered.

The duality between the sine-Gordon and the Thirring model has been proved for zero temperature~\cite{Coleman:1974bu} and finite temperature~\cite{Delepine:1997bz}.
It has been established that $\bar{\psi}\gamma ^{\mu}\psi$ corresponds to $\left(\beta/2\pi\right)\epsilon ^{\mu\nu}\partial _{\nu}\phi$, where $\phi$ is the scalar field in the sine-Gordon model.
Note that $\left(\beta/2\pi\right)\int dx \partial _x \phi$ is a topological charge measuring the winding number in $\phi$.
Consequently, the fermion number $\langle \bar{\psi}\gamma ^0 \psi\rangle$ corresponds to a topological charge in the sine-Gordon model.
Therefore, $\langle \bar{\psi}\gamma ^0 \psi\rangle$ serves as another key observable in this study.

\section{\label{sec:3}A brief introduction of the QMETTS algorithm}

QMETTS is a promising tool for studying the thermal properties of quantum systems, leveraging the concept of quantum typicality.
QMETTS computes the typical thermal states, which are pure states that serve as an efficient approximation for simulating the thermal properties of quantum systems.
The idea of QMETTS is to approximate $e^{-\beta H}|\phi \rangle$ using $e^{-ia^{-1}\beta A}|\phi\rangle$, where $A$ is an operator which will be introduced later. 
Then the trace ${\rm tr}\left[e^{-\beta H}\right]$ can be calculated using a Hilbert space $|\phi _i\rangle$ as ${\rm tr}\left[e^{-\beta H}\right]=\sum _i \langle \phi _i|e^{-\beta H}|\phi _i \rangle$.
Note that the trace can also be calculated in a stochastic approach using a set of randomly generated states if the Hilbert space is too large~\cite{Shen:2024fcj}.

We briefly introduce the QMETTS following Ref.~\cite{Czajka:2021yll}.
For an arbitrary state $\phi$, assume,
\begin{equation}
\begin{split}
&e^{-ia^{-1}\Delta \beta A}|\phi\rangle \approx \sqrt{C(\Delta \beta)} e^{-\Delta \beta H}|\phi\rangle,\\
\end{split}
\label{eq.3.1}
\end{equation}
when $\Delta \beta$ is small,
\begin{equation}
\begin{split}
&C(\Delta \beta)=\langle\phi| e^{-2 \Delta \beta H}|\phi\rangle \approx 1 - 2\Delta \beta \langle \phi |H | \phi \rangle,\\
\end{split}
\label{eq.3.2}
\end{equation}
where $\langle \phi |H | \phi \rangle$ can be obtained by measurement.

Denoting $j=l_0+l_1 \times 4 + l_2\times 4^2 + \ldots $ as a combined index of $\left(l_0,l_1,l_2,\ldots\right)$ with $l_i$ the integers satisfying $0\leq l_i \leq 3$, any Hermitian operator~(in our case the $A$ operator) on $N$ sites~(or $N$ qubits) can be expanded as,
\begin{equation}
\begin{split}
&A = \sum _j  c_j\hat{\sigma}_j=\sum _j  c_j\prod _{n}^N\sigma ^{l_n}(n),\\
\end{split}
\label{eq.3.3}
\end{equation}
where the $c_j$ are real coefficients, and $\sigma ^{0,1,2,3} = 1$, $\sigma ^x$, $\sigma ^t$, and $\sigma ^z$, respectively.
It is found that, Eq.~(\ref{eq.3.1}) provides a good approximation when $\Delta \beta$ is small, and $\{c_j\}$ is the solution of the equation,
\begin{equation}
\begin{split}
&\left ( S+S^{T}  \right ) c=b,
\end{split}
\label{eq.3.4}
\end{equation}
with,
\begin{equation}
\begin{split}
S_{j_1,j_2}&=\langle\phi| \hat{\sigma}_{j_i} \hat{\sigma}_{j_2}|\phi\rangle,\\
b_j&=\frac{-{\rm i}a}{\sqrt{C(\Delta \beta)}}\langle\phi|\left(H \hat{\sigma}_{j}-\hat{\sigma}_{j} H\right)|\phi\rangle,\\
\end{split}
\label{eq.3.5}
\end{equation}
where both $S$ and $b$ can be obtained by measurements or Hadamard tests.
The details of the measurements are shown in Appendix~\ref{sec:ap2}.

As will be introduced later, when considering an observable at $\beta$, it only needs to calculate $\phi(\beta/2)$. 
In this work, we divide $\beta/2$ by $K$ steps, i.e., $\Delta\beta = \beta / 2K$. 
Denoting $C(\Delta \beta, \beta')=1 - 2\Delta \beta \langle \phi(\beta') |H | \phi (\beta')\rangle$, and $A(c_j(\Delta \beta, \beta'))$ as the operator $A$ with coefficients $c_j$ solved by using $\phi(\beta')$, $\phi(\beta/2)$ can be obtained iteratively as,
\begin{equation}
\begin{split}
\left|\phi\left(\frac{\beta}{2}\right)\right\rangle &= \prod _{k=0}^{K-1} e^{-{\rm i}A\left(c_j(\Delta \beta, k\Delta \beta)\right)a^{-1}\Delta \beta}|\phi(0)\rangle,\\
C_{\phi} &= \prod _{k=0}^{K-1}C(\Delta \beta, k\Delta \beta),\\
\left|\Phi\left(\frac{\beta}{2}\right)\right\rangle &=\frac{1}{\sqrt{C_{\phi}}}\left|\phi\left(\frac{\beta}{2}\right)\right\rangle.
\end{split}
\label{eq.3.6}
\end{equation}
Each $e^{-{\rm i}A\left(c_j(\Delta \beta, k\Delta \beta)\right)a^{-1}\Delta \beta}|\phi(k\Delta \beta)\rangle$ can be evaluated using a Trotter decomposition with $t$ steps~\cite{Lloyd:1996aai}.

Then, for an observable $\hat{O}$ of interest,
\begin{equation}
\begin{split}
&\langle\hat{O}\rangle_{\beta}=\frac{{\rm tr}\left(e^{-\beta H} \hat{O}\right)}{{\rm tr}\left(e^{-\beta H}\right)},
\end{split}
\label{eq.3.7}
\end{equation}
with,
\begin{equation}
\begin{split}
{\rm tr}\left(e^{-\beta H} \hat{O}\right)&=\sum_i\left\langle \Phi_i \left(\frac{\beta}{2}\right)\right| \hat{O} \left|\Phi_i \left(\frac{\beta}{2}\right)\right\rangle,\\
{\rm tr}\left(e^{-\beta H}\right)&=\sum_i\left\langle \Phi_i \left(\frac{\beta}{2}\right)|\Phi_i \left(\frac{\beta}{2}\right)\right\rangle.
\end{split}
\label{eq.3.8}
\end{equation}
Then,
\begin{equation}
\begin{split}
&\langle\hat{O}\rangle_{\beta}=\frac{\sum _i C_{\phi _i}O_i}{\sum _i C_{\phi _i}},
\end{split}
\label{eq.3.9}
\end{equation}
with,
\begin{equation}
\begin{split}
&O_i = \left\langle \phi_i \left(\frac{\beta}{2}\right)\right| \hat{O} \left|\phi_i \left(\frac{\beta}{2}\right)\right\rangle,
\end{split}
\label{eq.3.10}
\end{equation}
which can be obtained by measurements.
The sum over $\phi _i$ is the sum over the Hilbert space $|\phi _i\rangle$.

\section{\label{sec:4}Numerical results}

In this work, $N=4$ is used. 
According to Eqs.~(\ref{eq.2.4}) and (\ref{eq.2.11}), the Hamiltonian can be expressed as, 
\begin{equation}
\begin{split}
aH_E &= h_{1}+amh_{2}-\frac{1}{4} g^{2}h_{3}-\frac{1}{4} g^{2}h_{4},\\
aH_M &= h_{1}+amh_{2}-\frac{1}{4} g^{2}h_{3}+\frac{1}{4} g^{2}h_{4},\\
\end{split}
\label{eq.4.1}
\end{equation}
with,
\begin{equation}
\begin{split}
h_{1}&=\frac{1}{2}\left \{ \sigma ^{x}\left ( 3 \right )\sigma ^{x}\left ( 2 \right )  +\sigma ^{x}\left ( 2 \right )\sigma ^{x}\left ( 1 \right )+\sigma ^{x}\left ( 1 \right )\sigma ^{x}\left ( 0 \right ) \right.\\
&\left.+\sigma ^{y}\left ( 3 \right )\sigma ^{y}\left ( 2 \right )+\sigma ^{y}\left ( 2 \right )\sigma ^{y}\left ( 1 \right )+\sigma ^{y}\left ( 1 \right )\sigma ^{y}\left ( 0 \right )\right.\\
&\left.+\sigma ^{x}\left ( 3 \right )\sigma ^{z}\left ( 2 \right )\sigma ^{z}\left ( 1 \right )\sigma ^{x}\left ( 0 \right )+\sigma ^{y}\left ( 3 \right )\sigma ^{z}\left ( 2 \right ) \sigma ^{z}\left ( 1 \right )\sigma ^{y}\left ( 0 \right )\right \},\\
h_{2}&=\frac{1}{2}\left \{ \sigma ^{z} \left ( 3 \right )-\sigma ^{z} \left ( 2 \right )+ \sigma ^{z} \left ( 1 \right ) - \sigma ^{z} \left ( 0 \right )\right \},\\
h_{3}&=\left \{ \sigma ^{z} \left ( 3 \right )+\sigma ^{z} \left ( 2 \right )+ \sigma ^{z} \left ( 1 \right ) + \sigma ^{z} \left ( 0 \right )\right \}\\
&+\frac{1}{4}\left \{ \sigma ^{z} \left ( 3 \right )\sigma ^{z} \left ( 2 \right )+\sigma ^{z} \left ( 2 \right )\sigma ^{z} \left ( 1 \right )+\sigma ^{z} \left ( 1 \right )\sigma ^{z} \left ( 0 \right )
+\sigma ^{z} \left ( 3 \right )\sigma ^{z} \left ( 0 \right )\right \}, \\
h_{4}&=\frac{1}{4}\left \{\sigma ^{z} \left ( 3 \right )\sigma ^{z} \left ( 2 \right )+\sigma ^{z} \left ( 2 \right )\sigma ^{z} \left ( 1 \right )+\sigma ^{z} \left ( 1 \right )\sigma ^{z} \left ( 0 \right )
+\sigma ^{z} \left ( 3 \right )\sigma ^{z} \left ( 0 \right )\right \}.
\end{split}
\label{eq.4.2}
\end{equation}

According to Eq.~(\ref{eq.2.11}), the observables of interest are,
\begin{equation}
\begin{split}
\langle \bar{\psi}\psi \rangle &= \langle h_2\rangle,\\
\langle \bar{\psi}\gamma ^0\psi \rangle &= 2+\frac{1}{2}\langle \sigma^z(0)+\sigma^z(1)+\sigma^z(2)+\sigma^z(3)\rangle.\\
\end{split}
\label{eq.4.3}
\end{equation}

In this work, the mass is fixed at $m=0.5a^{-1}$.
For $H_E$, the range of $0.01a^{-1}\leq T \leq\;a^{-1}$ and $0\leq g^2\leq 3$ is considered, where $T=1/\beta$.
For $H_M$, the range of $0.01a^{-1}\leq T \leq\;a^{-1}$ and $0\leq g^2\leq 5$ is considered.
When $a^{-1}=100\;{\rm MeV}$, the parameters are $m=50\;{\rm MeV}$ and $1\;{\rm MeV}\leq T \leq 100\;{\rm MeV}$.

\begin{figure}[htbp]
\includegraphics[width=0.48\hsize]{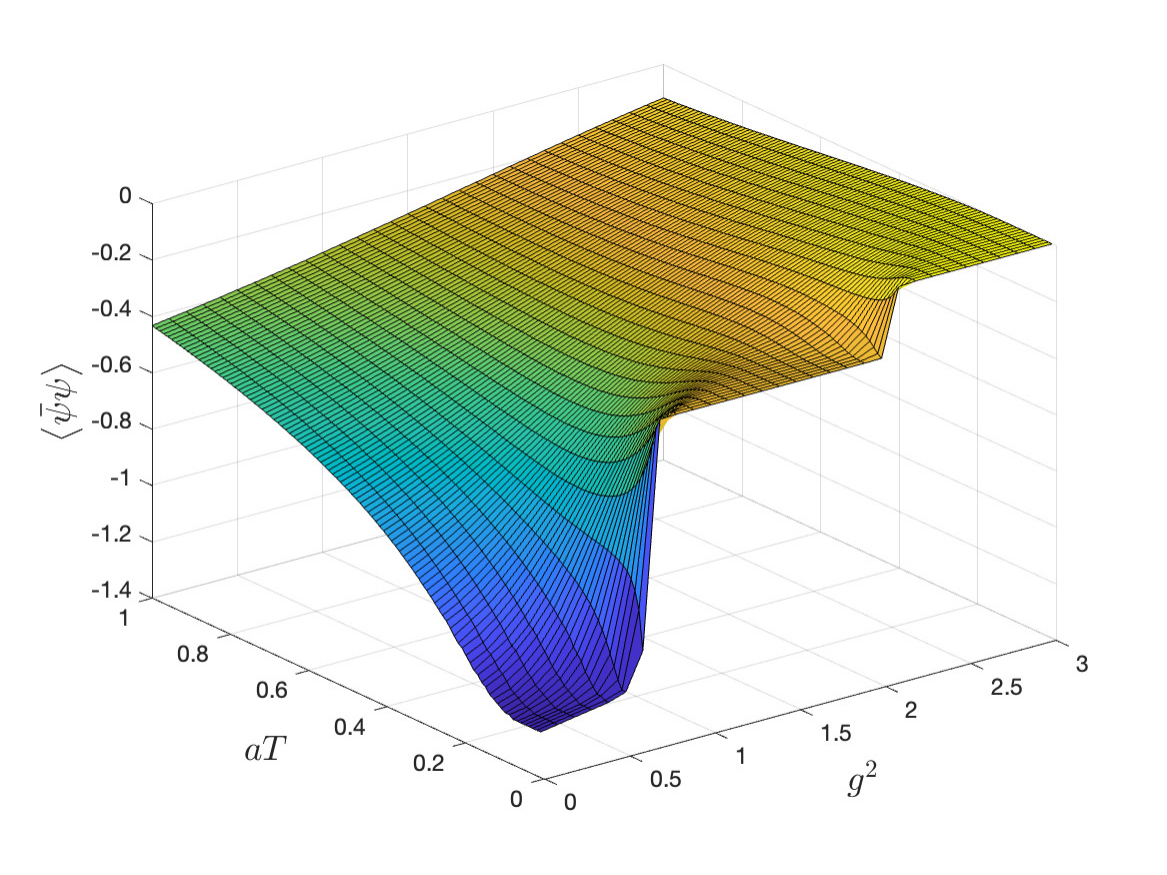}
\includegraphics[width=0.48\hsize]{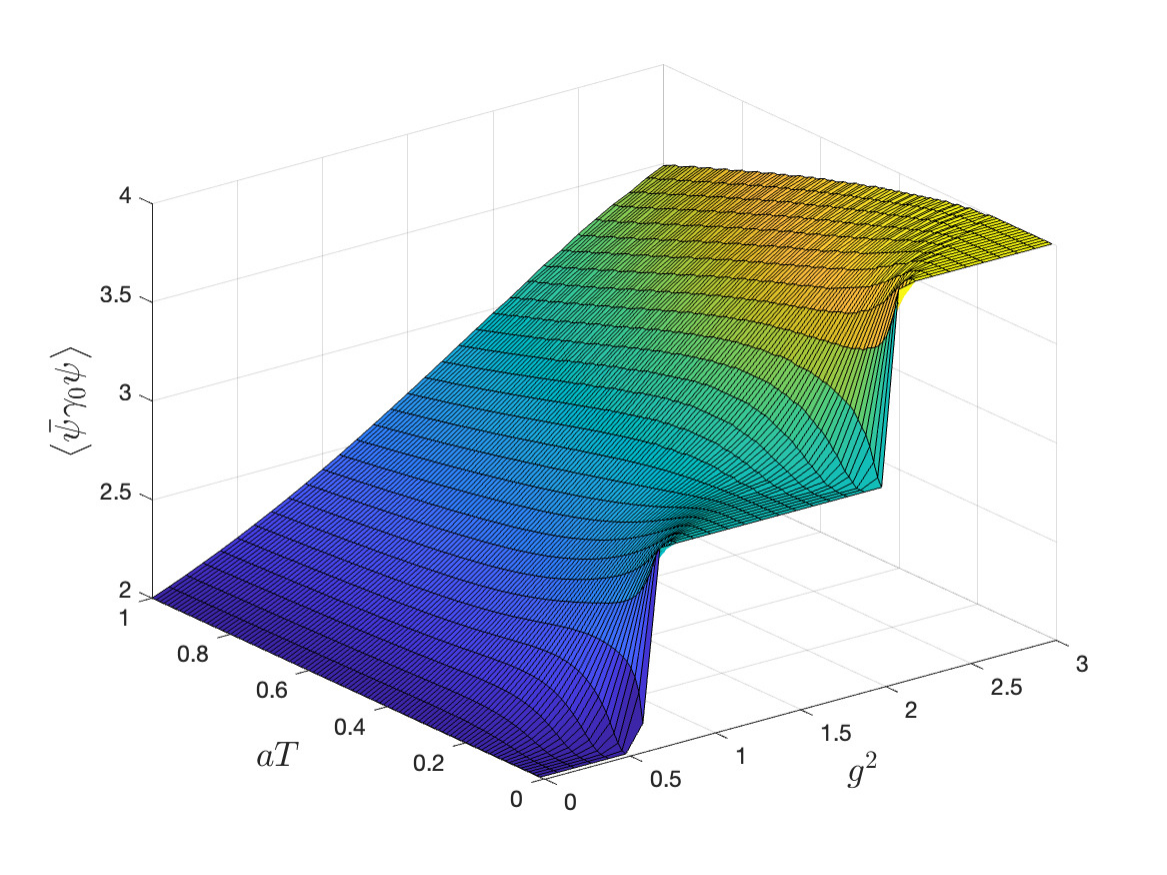}\\
\includegraphics[width=0.48\hsize]{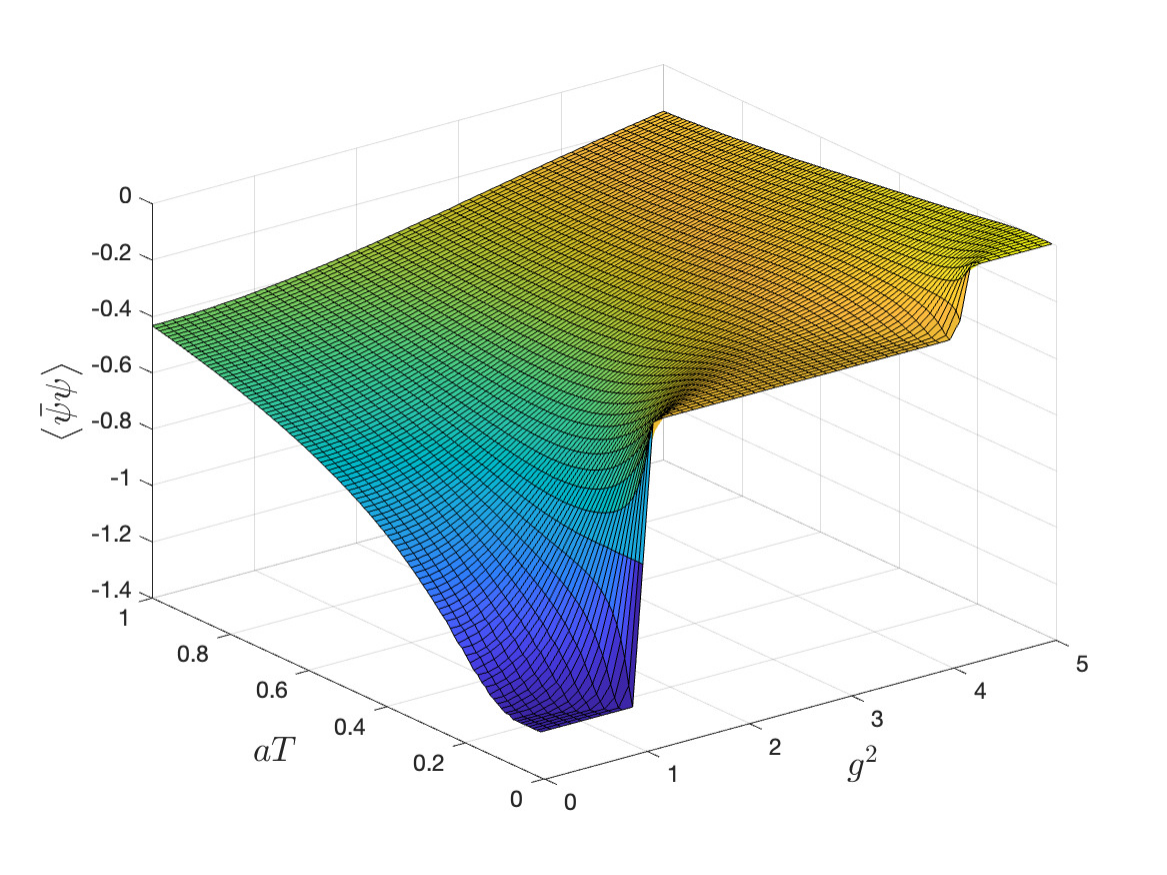}
\includegraphics[width=0.48\hsize]{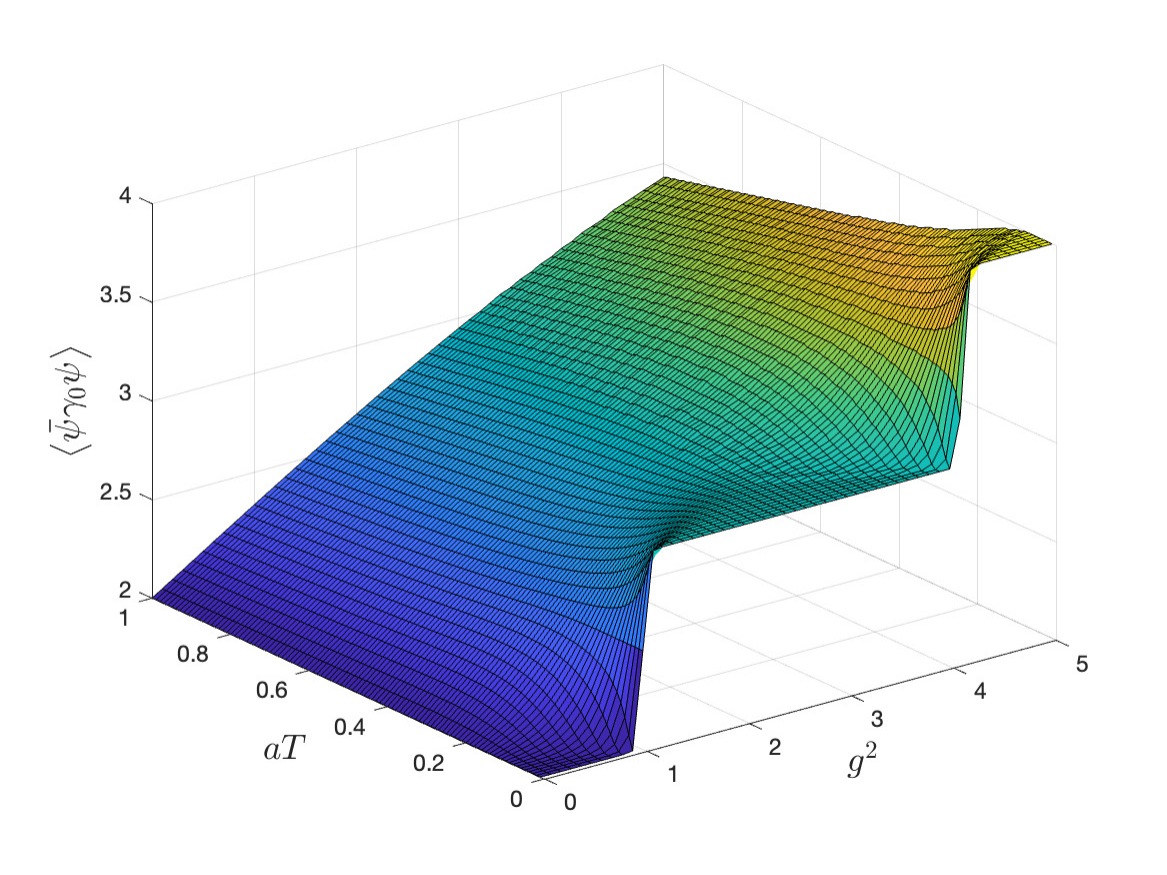}
\caption{\label{fig:exactdiagonal}Chiral condensation $\langle \bar{\psi}\psi\rangle$~(the left panels) and fermion number~(topological charge) $\langle \bar{\psi}\gamma^0\psi\rangle$~(the right panels) as functions of $T$ and $g^2$ calculated by using exact diagonalization. 
The panels in the first row correspond to $H_E$, and the ones in the second row correspond to $H_M$, respectively.
Chiral symmetry breaks at low temperatures or weak coupling strengths.
At low temperatures, the topological charge takes integer values, indicating a topological structure.}
\end{figure}
Since $N=4$, exact diagonalization can be used to compute $e^{-\beta H}$.
To focus on QMETTS numerical results, discussion of discretization errors is deferred to the next section.
The observables $\langle \bar{\psi}\psi\rangle$ and $\langle \bar{\psi}\gamma^0\psi\rangle$ are calculated using the Hamiltonian in Eq.~(\ref{eq.4.1}).
The results of $\langle \bar{\psi}\psi\rangle$ and $\langle \bar{\psi}\gamma^0\psi\rangle$ using exact diagonalization are shown in Fig.~\ref{fig:exactdiagonal}.
It can be seen that, the chiral symmetry is broken at either low temperatures or weak couplings.
At low temperatures, the dependence of chiral condensation on the coupling constant is manifest in the form of distinct plateaus, which correspond to different topological charges.
The values of the topological charge are integers at low temperatures, therefore exhibiting a topological structure.

The goal of our study is to reproduce the above results using QMETTS.
Since $N=4$ is small, we use a complete set of bases of the Hilbert space, $\Phi _i = |i\rangle$, for $0\leq i \leq 15$.
The range of temperature is $0.1a^{-1}\leq T \leq 2a^{-1}$, in other words, $0.5\leq a^{-1}\beta \leq 10$ with $a^{-1}\Delta \beta = 0.25$ and $K=20$, where $K=\beta _{max}/(2\Delta \beta)$ is defined in previous section.
Therefore, the temperatures in consideration is $1/\left(2k\Delta \beta\right)$ with $1\leq k\leq K$.
The number of steps in Trotter decomposition is $t=10$.
In the evaluation, only the $\hat{\sigma}_j$ terms in Eq.~(\ref{eq.3.3}) with $|c_j|>0.001$ are kept.
In the case of $H_M$, the values of the coupling constant are chosen as $g^2=0,0.2,0.4,\ldots ,5.0$.
In the case of $H_E$, the values of the coupling constant are chosen as $g^2=0,0.1,0.2,\ldots ,3.0$.

\begin{figure}[htbp]
\includegraphics[width=0.48\hsize]{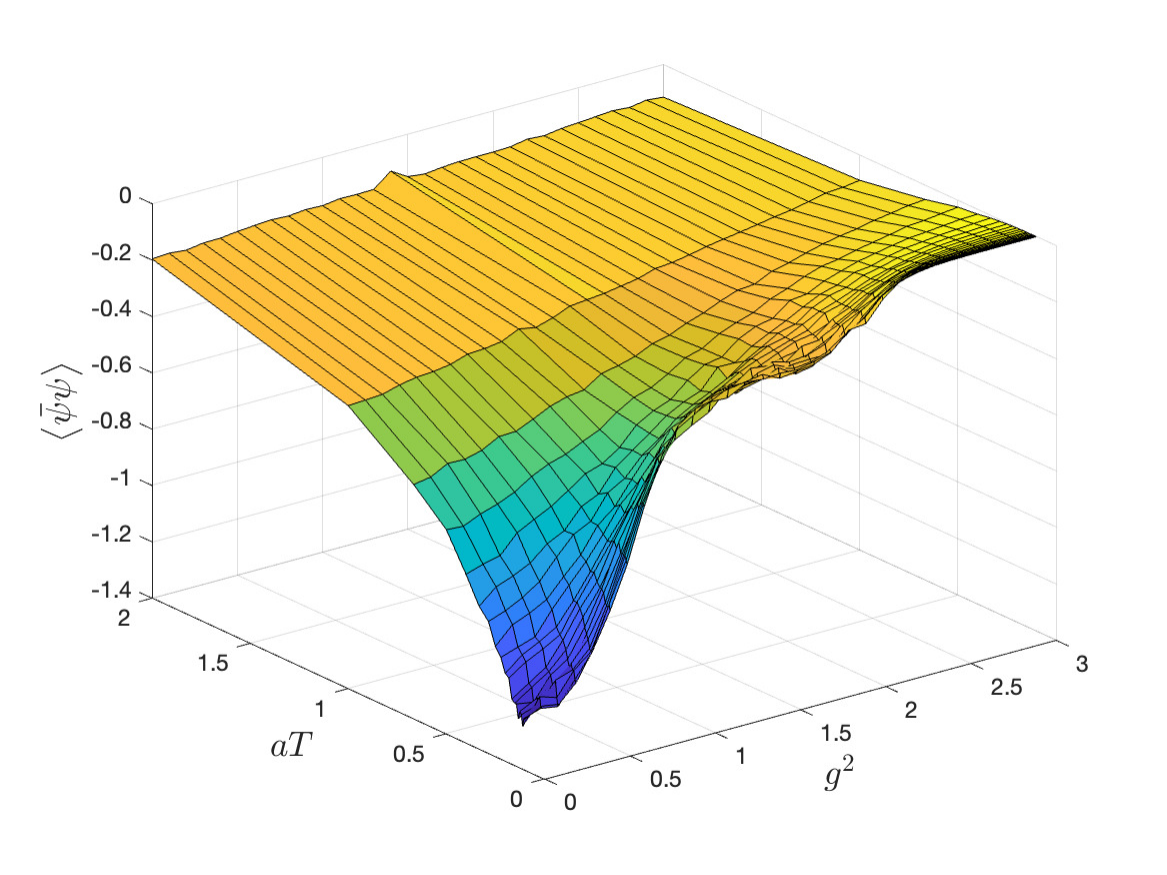}
\includegraphics[width=0.48\hsize]{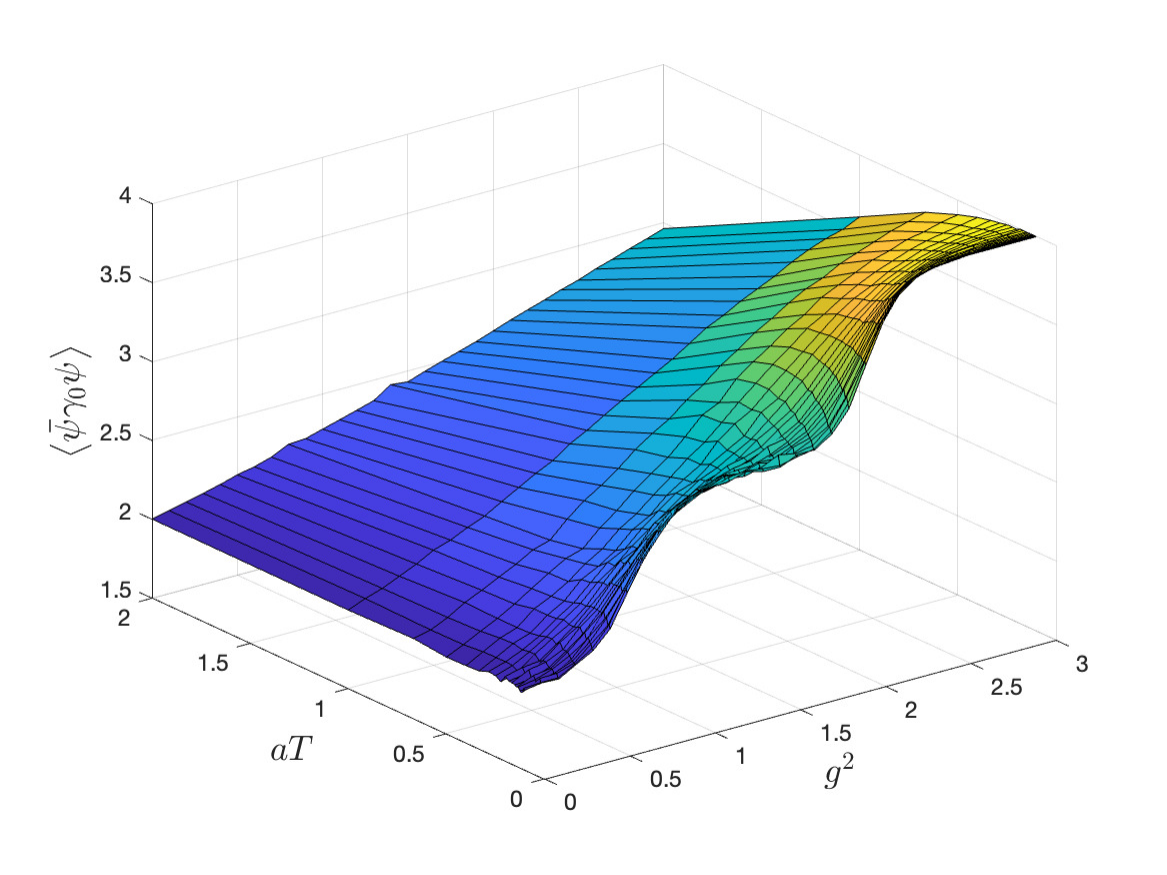}\\
\includegraphics[width=0.48\hsize]{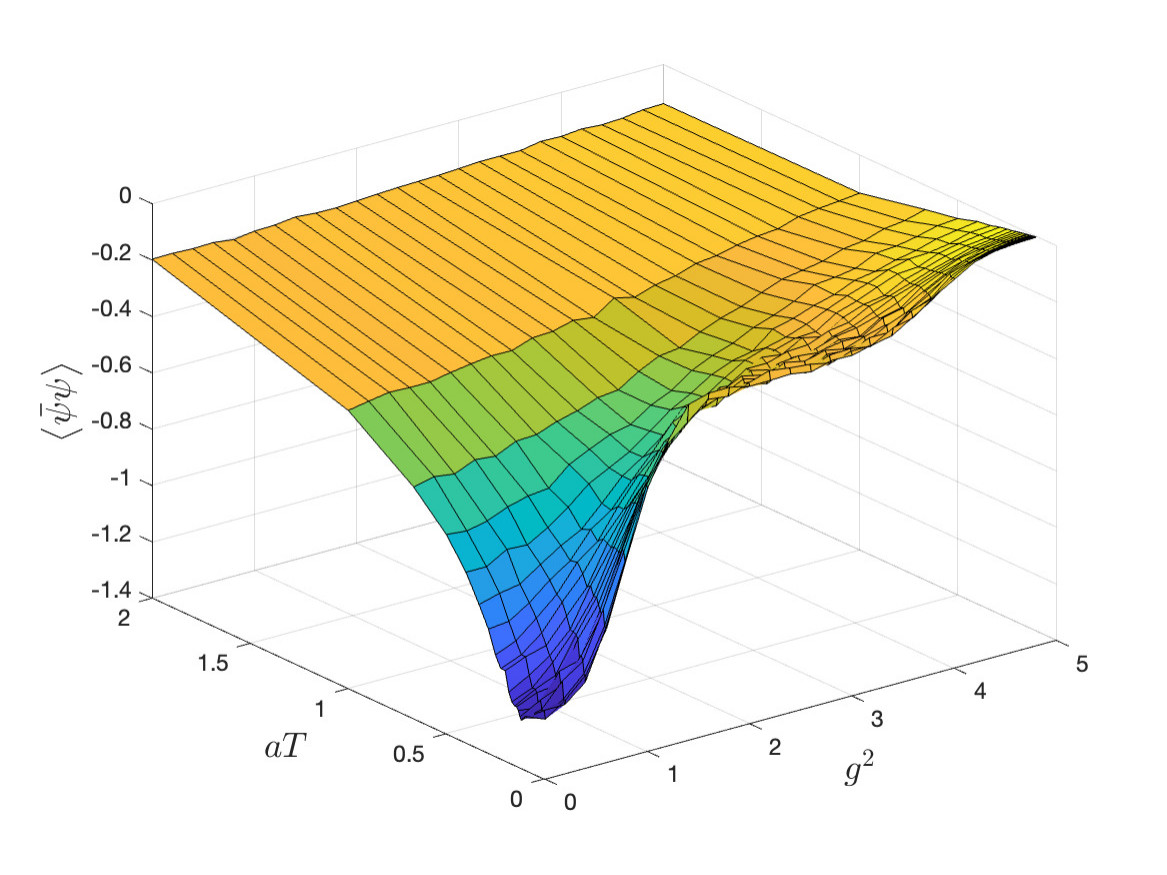}
\includegraphics[width=0.48\hsize]{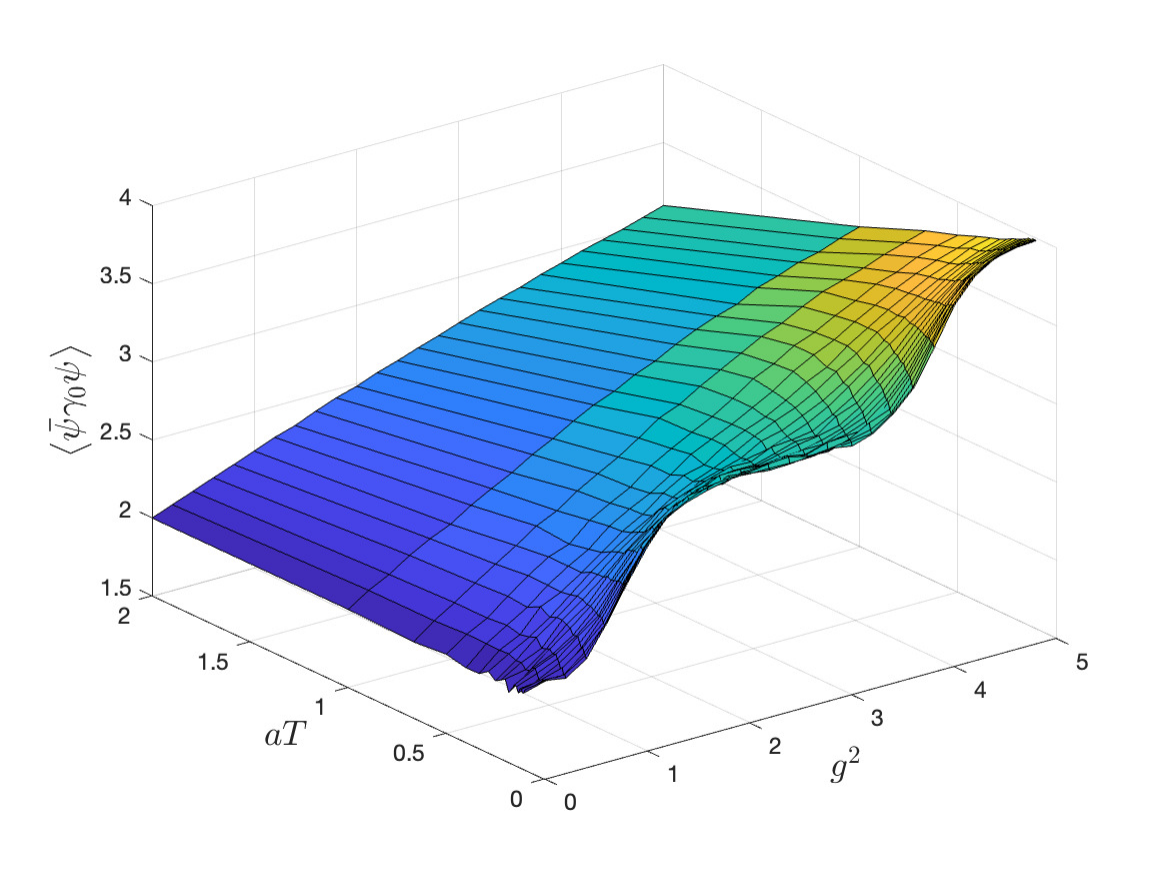}
\caption{\label{fig:quantumresult}$\langle \bar{\psi}\psi\rangle$~(the left panels) and $\langle \bar{\psi}\gamma^0\psi\rangle$~(the right panels) as functions of $T$ and $g^2$ calculated by using QMETTS. 
The panels in the first row correspond to $H_E$, and the ones in the second row correspond to $H_M$, respectively.
For clarity, statistical uncertainties are omitted , with their analysis deferred to the following section. 
Comparison with Fig.~\ref{fig:exactdiagonal} demonstrates that QMETTS generally captures the evolution of both the chiral condensation and topological charge with varying temperature and interaction strength.}
\end{figure}
The circuit is implemented using \verb"qiskit"~\cite{qiskit2024}, and the results are obtained by using a simulator on a classical computer.
The observables $\langle \bar{\psi}\psi\rangle$ and $\langle \bar{\psi}\gamma^0\psi\rangle$ at different temperatures and coupling constants are shown in Fig.~\ref{fig:quantumresult}.
For simplicity, the parameters to calculate $c_j$ and $C(\beta)$ used in the evaluation are measured exactly which is only possible using a simulator, while $O_i$ in Eq.~(\ref{eq.3.10}) are measured for $r=1024$ times.
As can be seen from Figs.~\ref{fig:exactdiagonal} and \ref{fig:quantumresult}, the outcomes yielded by QMETTS are largely in alignment with those obtained through exact diagonalization. 
Among these results, the breaking and restoration of chiral symmetry, the plateaus in chiral condensation, and the behavior of topological charges are all consistent. 
In QMETTS, the jitter in the values is primarily attributable to the number of measurements.
The edges of the plateaus at low temperatures are observed to be less sharp than in exact diagonalization, due to the fact that in QMETTS we only evolve up to $a^{-1}\beta=10$, in contrast to the exact diagonalization where $a^{-1}\beta=100$.

In general, QMETTS reproduces results that are in accordance with those obtained through exact diagonalization. 
The outcomes demonstrate the breaking and restoration of chiral symmetry at different temperatures and different coupling constants. 
Furthermore, the integer values of the topological charges at low temperatures and the plateaus where the chiral symmetry varies with the coupling constant are also established.

\section{\label{sec:5}A discussion on the errors}

There are three primary sources of errors in quantum simulations of the Thirring model, the model-related errors, algorithmic errors, and implementation errors.
Due to the substantial circuit depth required for low-temperature simulations, which currently prevents physical implementation, this section focuses on model-related and algorithmic errors.
In this work, we neglect noise effects to demonstrate that the QITE algorithm can inherently handle low-temperature systems with topological phase transitions under ideal conditions. 
The study of noise resilience is deferred to future research.

\subsection{\label{sec:5.1}The systematic errors of the model}

By applying Jordan-Wigner transformations, the Hamiltonians of the Thirring model are mapped to those in Eq.~(\ref{eq.4.1}).
However, Hamiltonians in Eq.~(\ref{eq.4.1}) are subject to both discretization errors and finite-volume effects.

\begin{figure}[htbp]
\includegraphics[width=0.48\hsize]{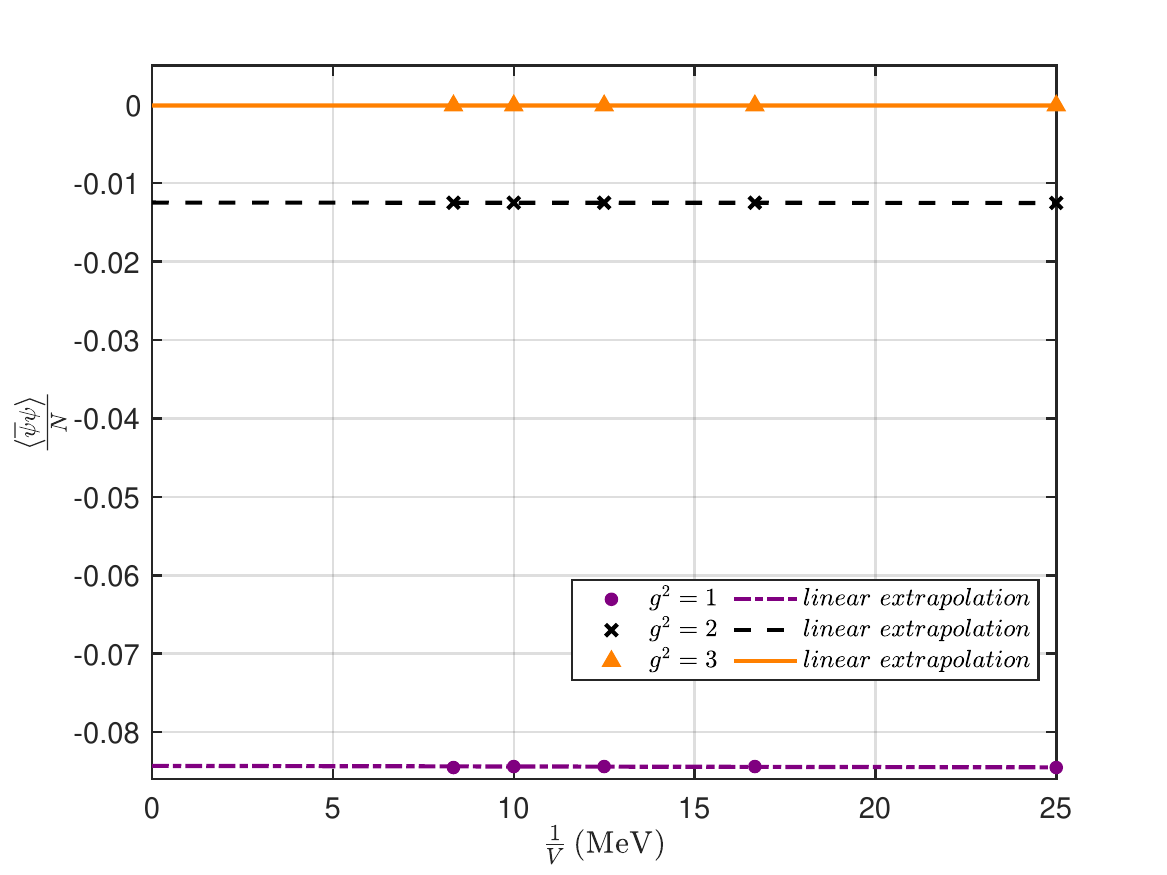}
\includegraphics[width=0.48\hsize]{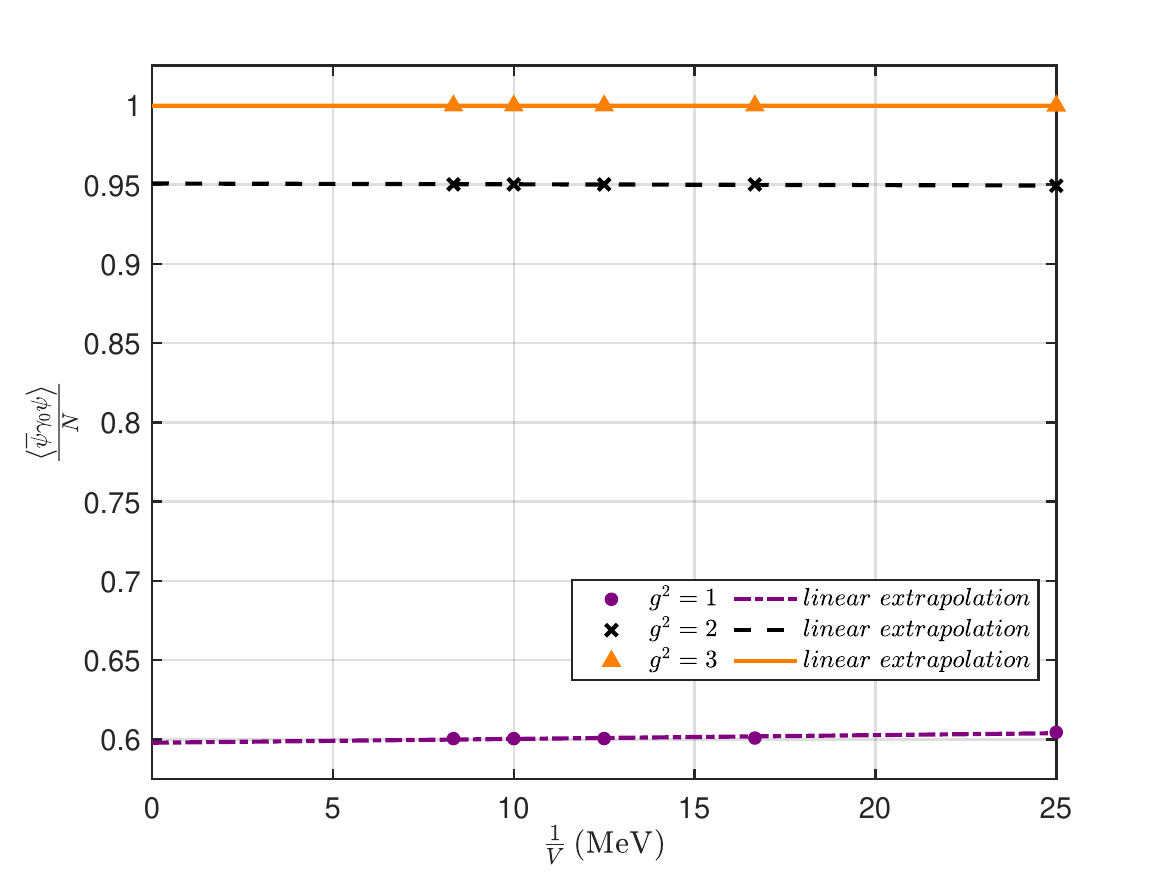}\\
\includegraphics[width=0.48\hsize]{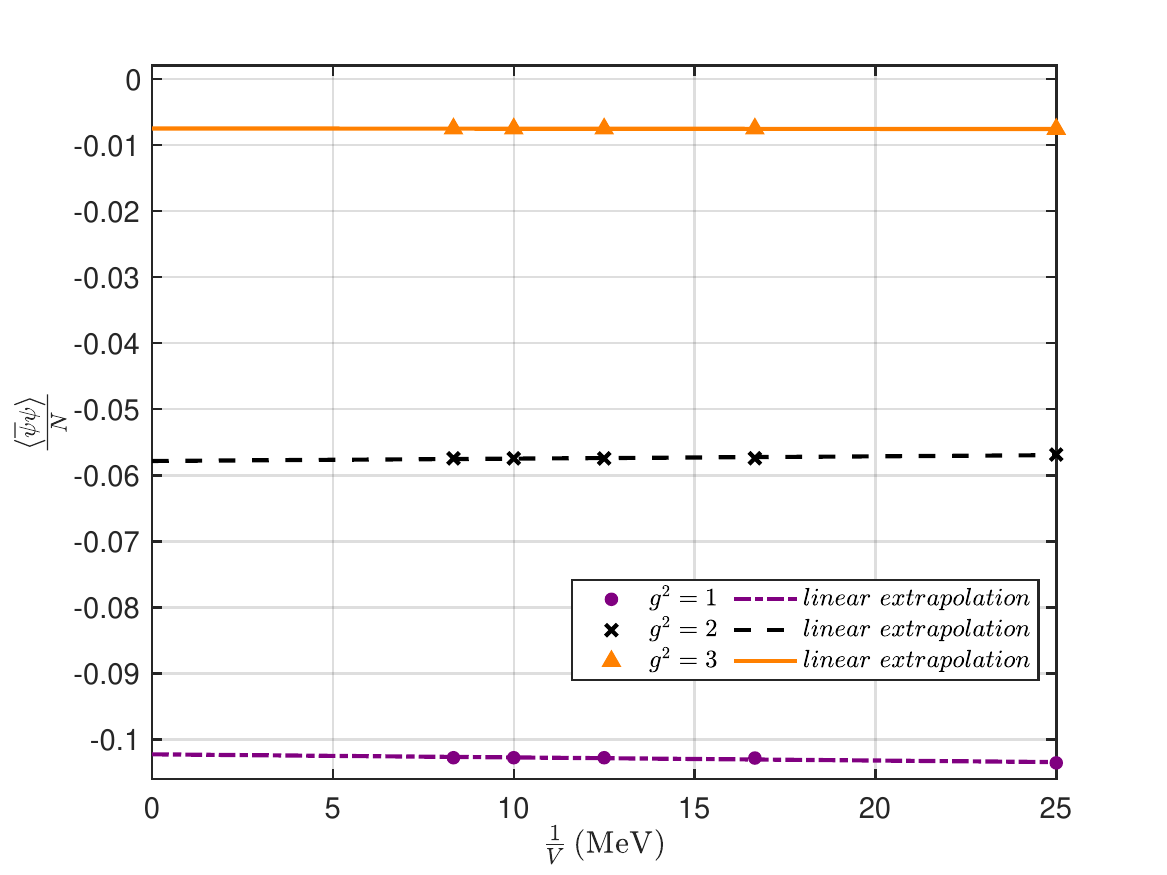}
\includegraphics[width=0.48\hsize]{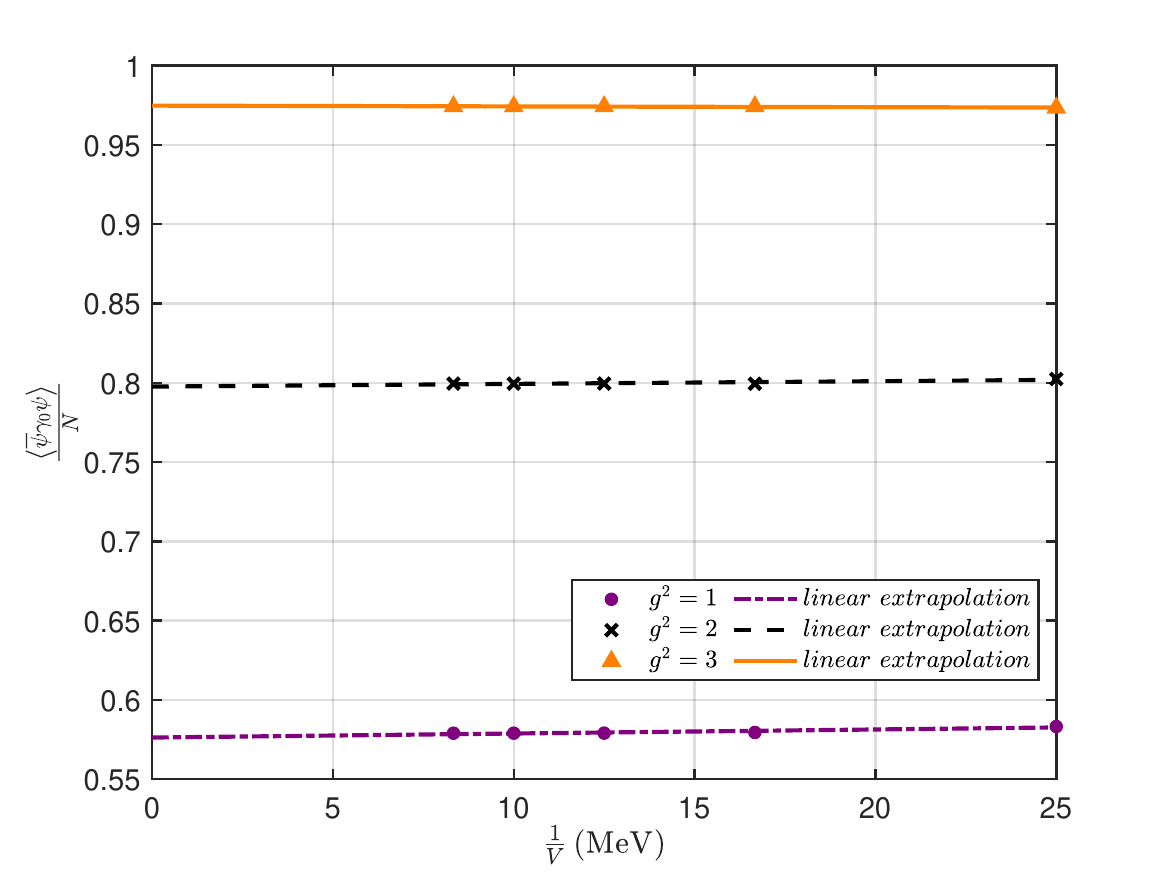}
\caption{\label{fig:fv-beta1}
At $a^{-1}\beta=1$~($T=100\;{\rm MeV}$) and different $g^2$, the average chiral condensate $\langle \bar{\psi}\psi\rangle / N$ and topological charge $\langle \bar{\psi}\gamma _0\psi\rangle / N$ as functions of $1/V$ for lattice sizes $N=4$, $6$, $8$, $10$, and $12$, along with corresponding linear extrapolations.
The upper panels are the results of $H_M$, the bottom panels are results of $H_E$, respectively.
It can be observed that finite-volume effects are negligible at higher temperatures.}
\end{figure}
\begin{figure}[htbp]
\includegraphics[width=0.48\hsize]{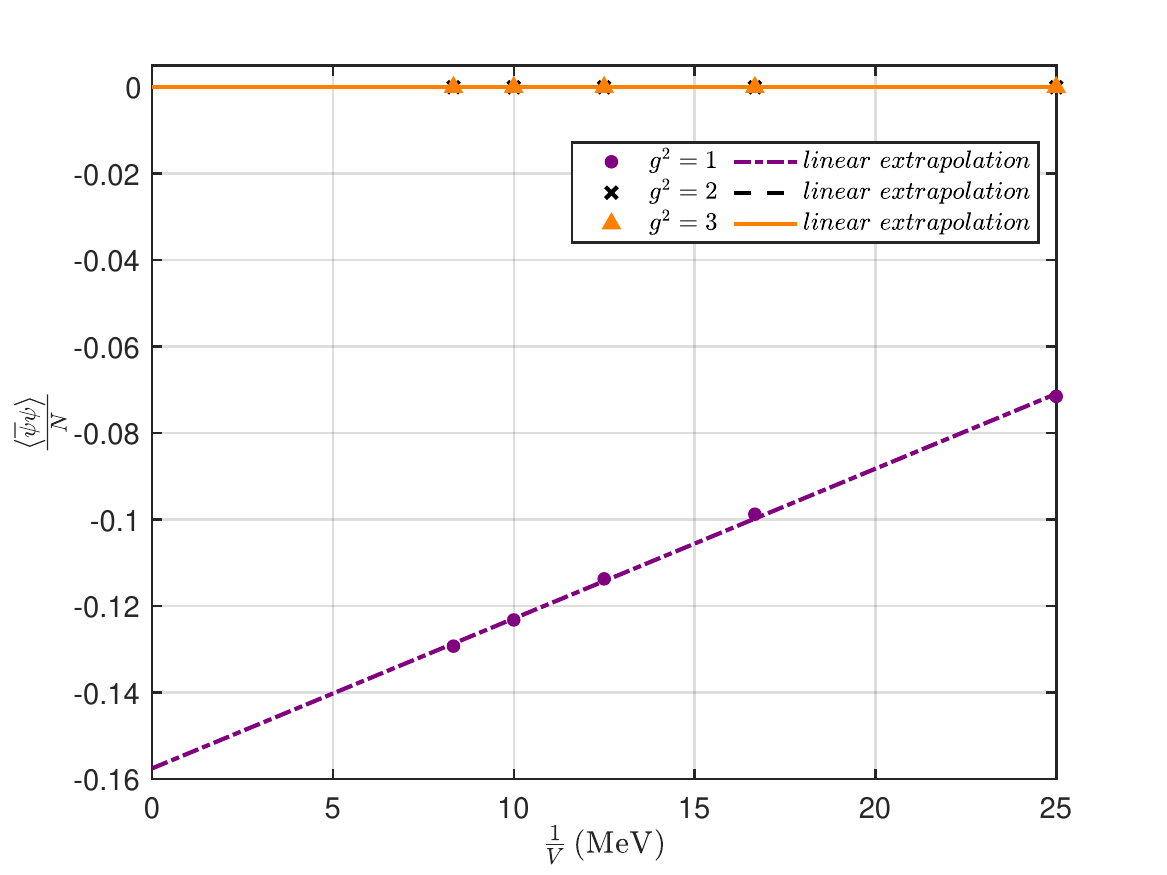}
\includegraphics[width=0.48\hsize]{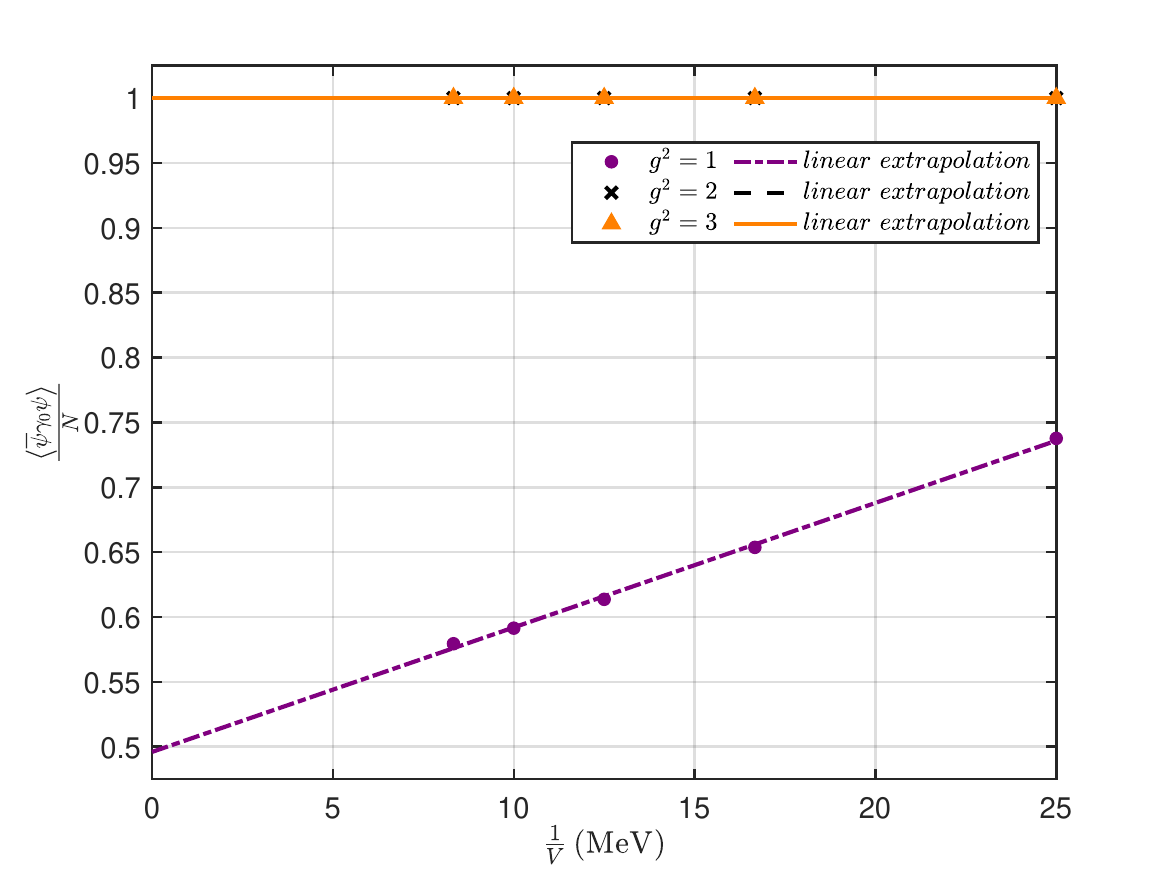}\\
\includegraphics[width=0.48\hsize]{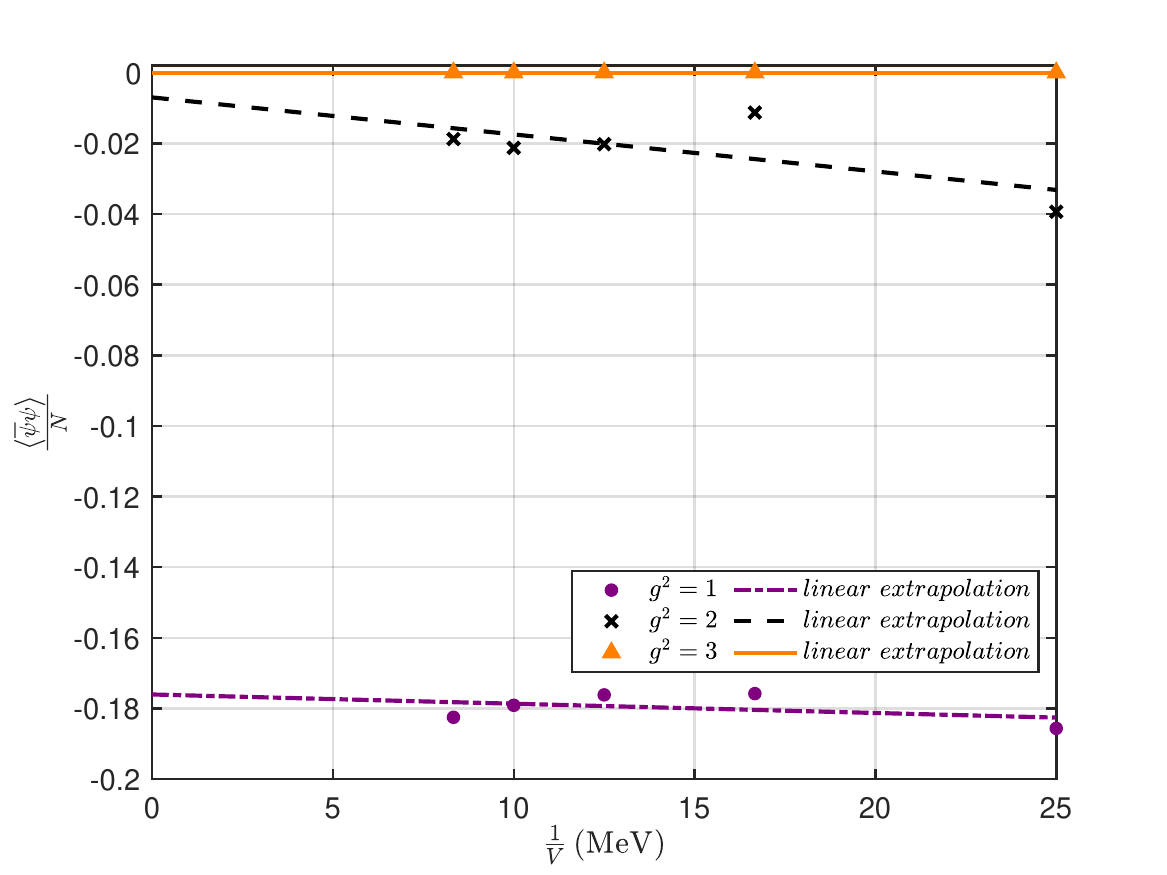}
\includegraphics[width=0.48\hsize]{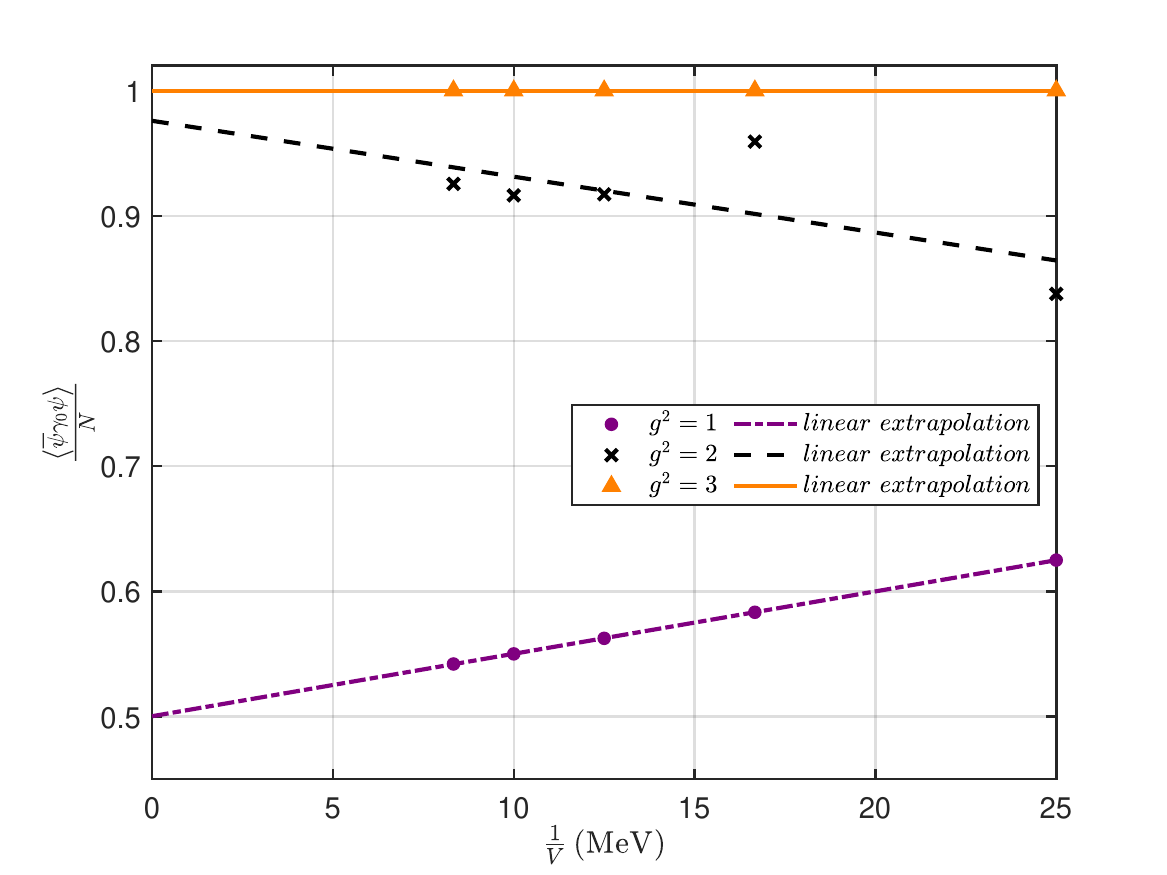}
\caption{\label{fig:fv-beta10}Same as Fig.~\ref{fig:fv-beta1} but for $a^{-1}\beta=10$~($T=10\;{\rm MeV}$).
The effect of finite volume is significant at smaller $g^2$, especially for the case of $H_M$~(the upper panels).
However, the regime with $1/V \to 0$ can be effectively captured through linear extrapolations.
}
\end{figure}
To analyze finite volume effects, we examine the asymptotic behavior of observables via exact diagonalization across increasing system sizes.
In order to consider only the influence of the finite volume, we calculate the average of chiral condensation and the average of topological charge, i.e., $\langle \bar{\psi}\psi\rangle/N$ and $\langle \bar{\psi}\gamma _0\psi\rangle/N$.
The cases of $N=4$, $6$, $8$, $10$ and $12$ are studied, the results at $a^{-1}\beta=1$ and $10$ for $g^2=1$, $2$ and $3$ are shown in Figs.~\ref{fig:fv-beta1} and \ref{fig:fv-beta10}.

The results are shown as functions of $1/V$ where $V=aN$ is the volume.
It can be found that, at higher temperatures~($a^{-1}\beta=1$), the effect of finite volume is minor, and even results at $N=4$ closely match those at $N=12$. 
In contrast, at lower temperatures~($a^{-1}\beta=10$), the effect of finite volume becomes significant for smaller $g^2$, particularly for the case of $H_M$.
This is understandable because when topological order arises, the larger the volume, the more difficult it is for the system to switch between different topological sectors.
When $g^2$ is relatively large, the topological sector of the final state is closer to the initial state, whereas when $g^2$ is relatively small, the transition from the initial state to the final state requires switching between different topological sectors, explaining the observed sensitivity.
In Figs.~\ref{fig:fv-beta1} and \ref{fig:fv-beta10}, we simultaneously present results from linear extrapolation. 
It can be observed that although the finite volume effect has significant impact, the worst case scenario is the chiral condensate at $g^2=1$ and $a^{-1}\beta=10$, where for $N=4$ $\langle \bar{\psi}\psi\rangle/N\approx -0.07$ and for $N=12$ one has $\langle \bar{\psi}\psi\rangle/N\approx -0.13$, the regime with $1/V \to 0$ can be effectively captured through linear extrapolations.

\begin{figure}[htbp]
\includegraphics[width=0.48\hsize]{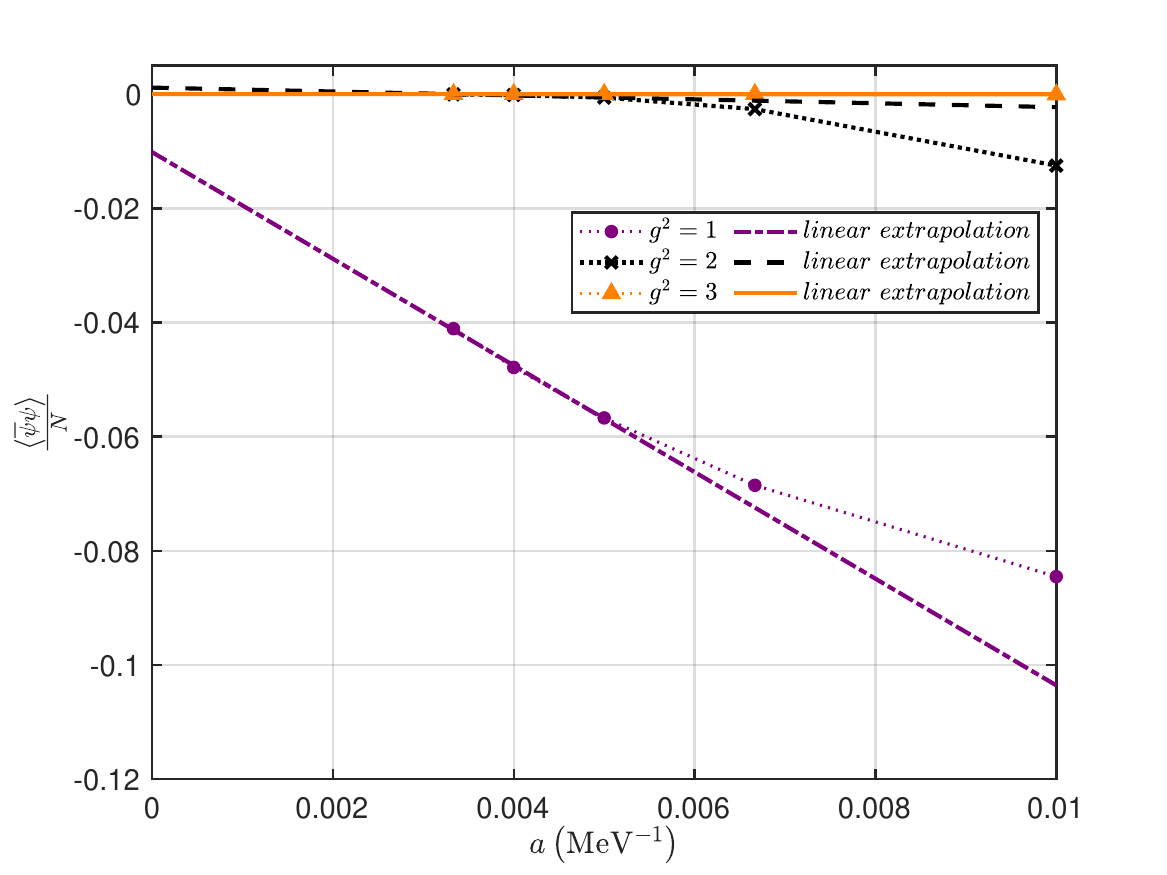}
\includegraphics[width=0.48\hsize]{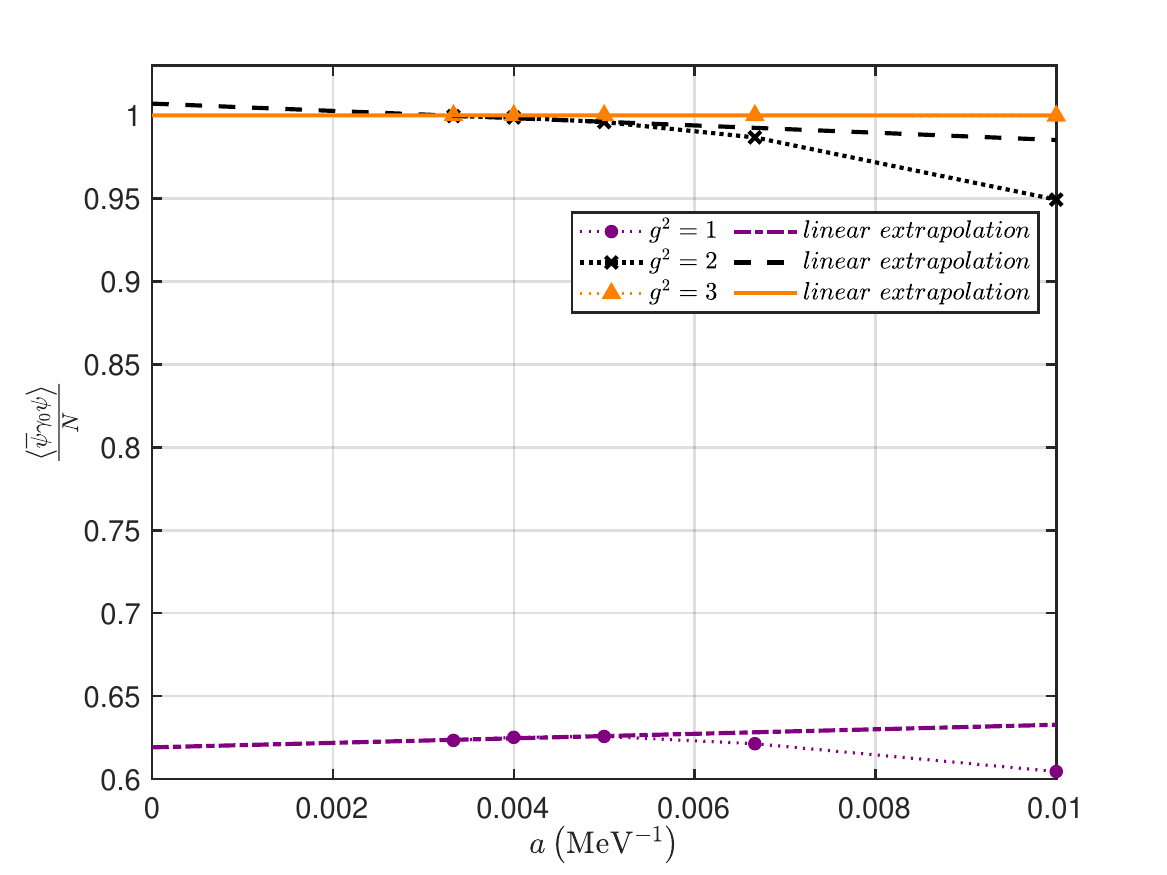}\\
\includegraphics[width=0.48\hsize]{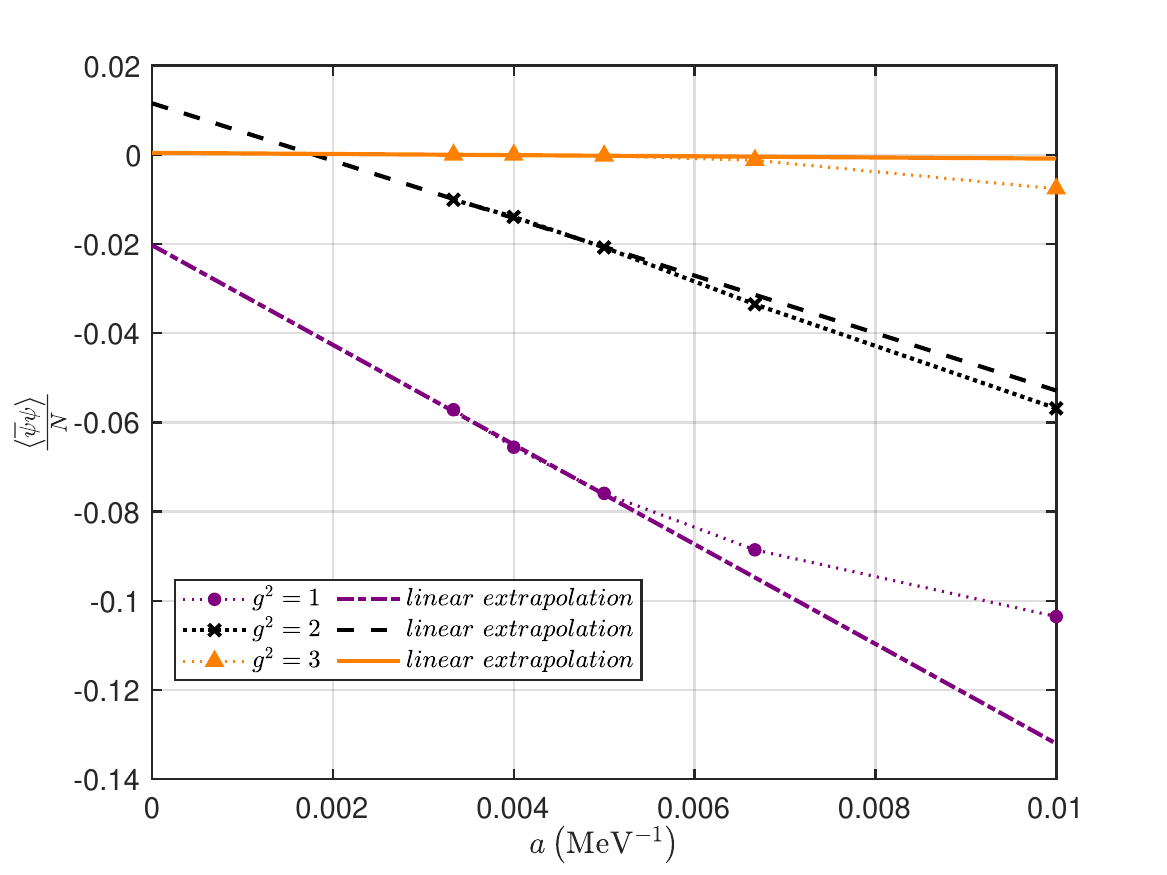}
\includegraphics[width=0.48\hsize]{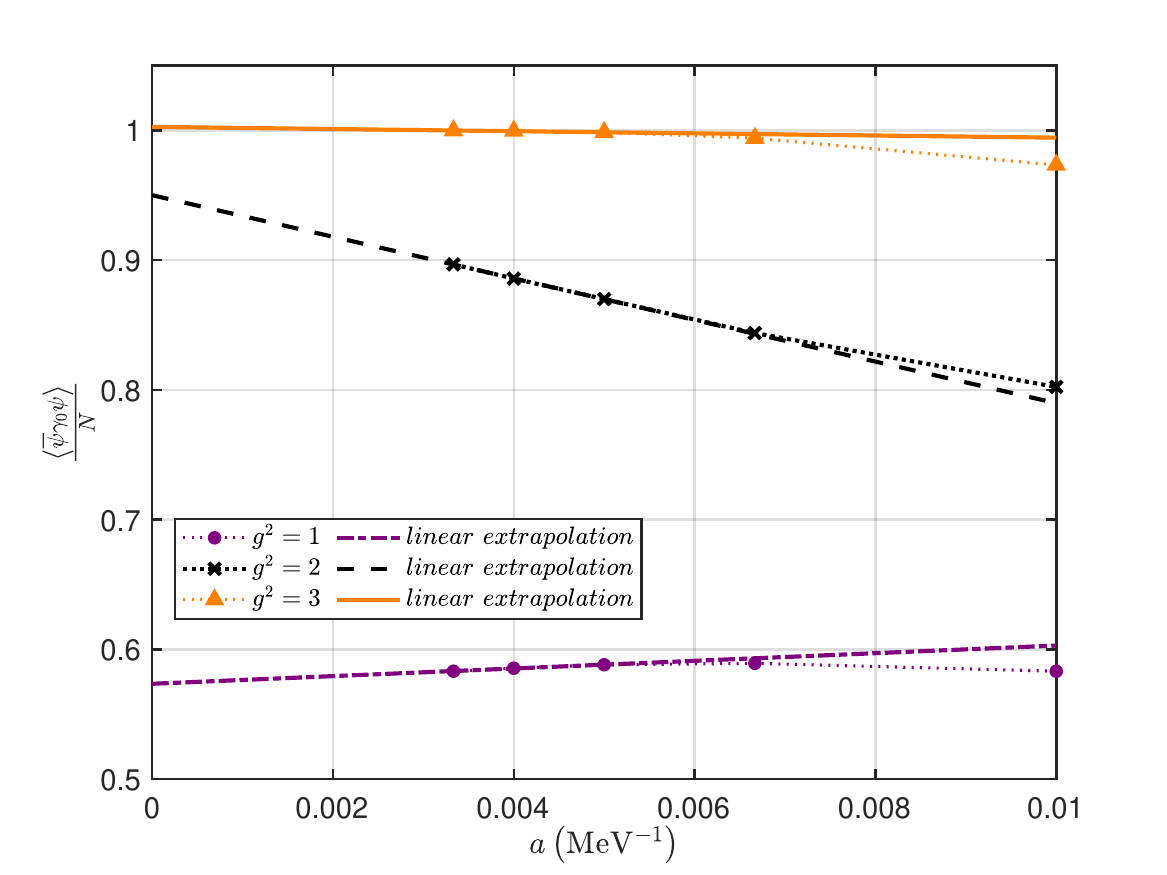}
\caption{\label{fig:de-beta1}.At $T=100\;{\rm MeV}$ and different $g^2$, the average chiral condensate $\langle \bar{\psi}\psi\rangle / N$ and topological charge $\langle \bar{\psi}\gamma _0\psi\rangle / N$ as functions of $a$ for lattice sizes $N=4$, $6$, $8$, $10$, and $12$, along with corresponding linear extrapolations.
The upper panels are the results of $H_M$, the bottom panels are results of $H_E$, respectively.
The convergence behavior of observables is not purely linear, therefore the linear extrapolation is performed with the data points of $N=8,10,12$.}
\end{figure}
\begin{figure}[htbp]
\includegraphics[width=0.48\hsize]{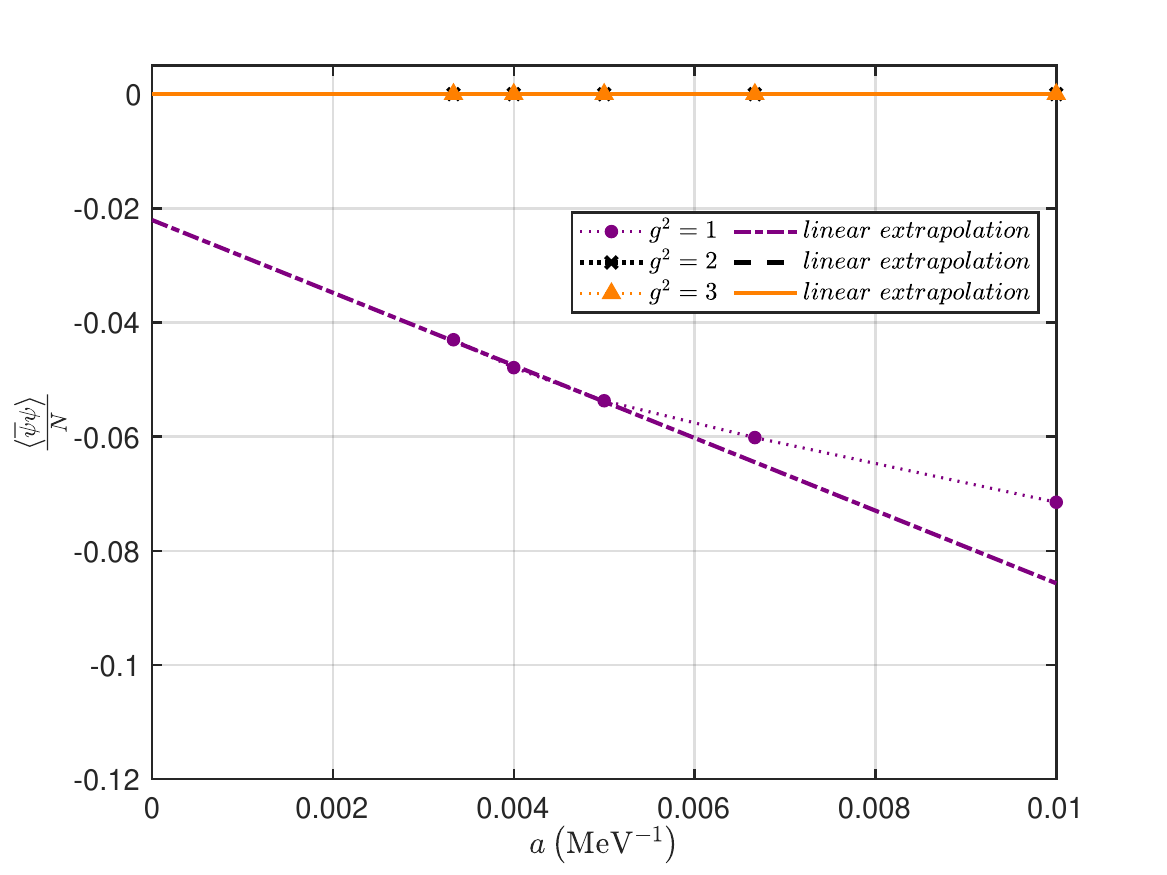}
\includegraphics[width=0.48\hsize]{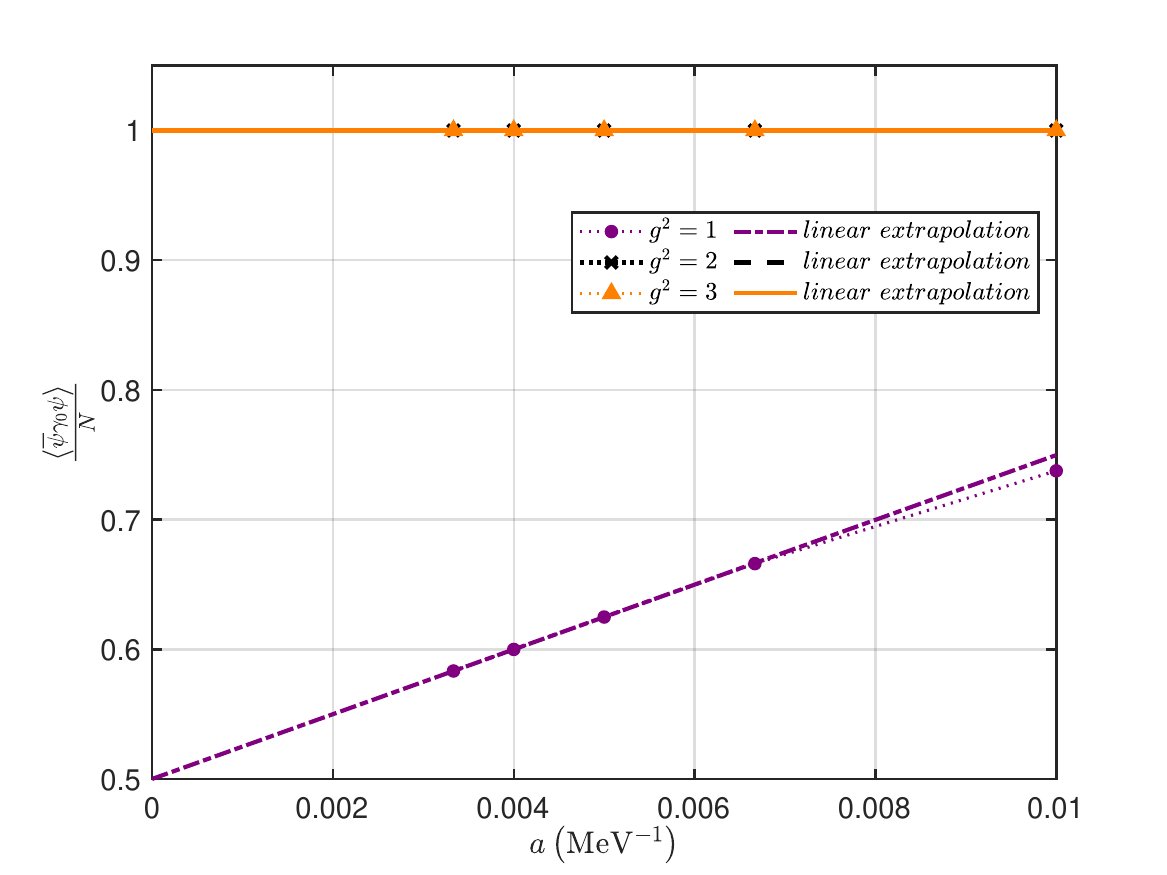}\\
\includegraphics[width=0.48\hsize]{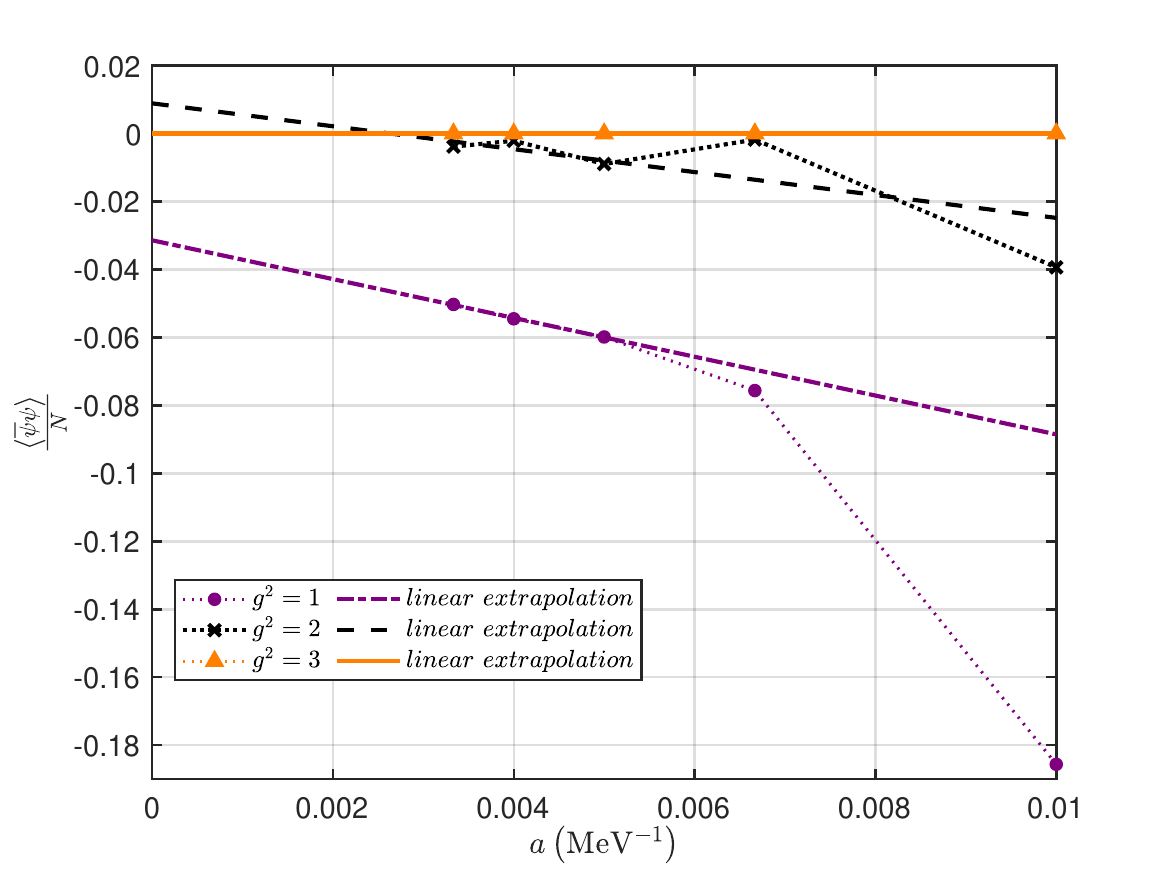}
\includegraphics[width=0.48\hsize]{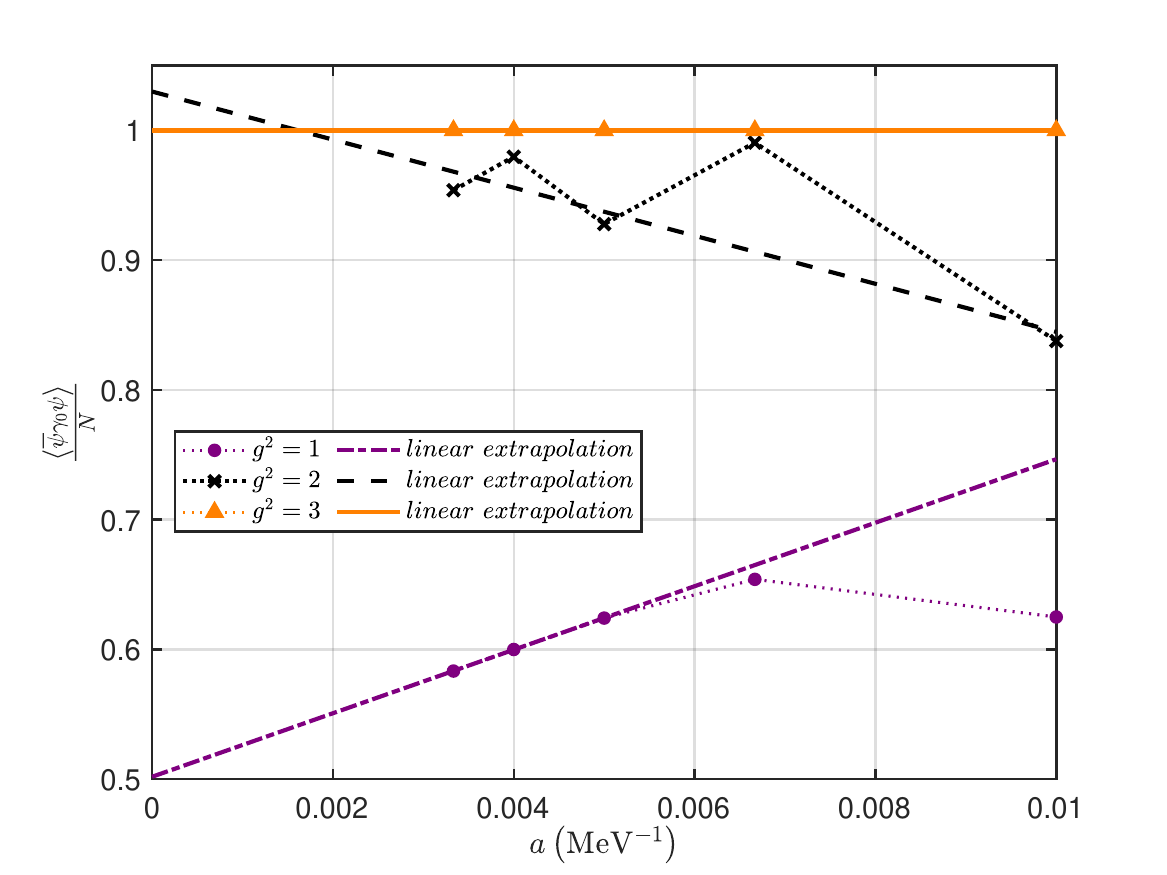}
\caption{\label{fig:de-beta10}.Same as Fig.~\ref{fig:de-beta1} but for $T=10\;{\rm MeV}$.
For chiral condensation, the slope of the linear extrapolation is smaller than those at $T=100\;{\rm MeV}$.
For the $H_E$~(the bottom panels), convergence exhibits oscillatory behavior at certain $g^2$ values. 
To obtain more reliable results, additional data points are required for linear extrapolation.}
\end{figure}
Similarly the discretization error is also studied by using $N=4$, $6$, $8$, $10$ and $12$ lattices but with $a^{-1}=100$, $150$, $200$, $250$ and $300\;{\rm MeV}$.
To make sure the model is the same one under different lattice spacings, we consider the cases of $T=100\;{\rm MeV}$ and $10\;{\rm MeV}$ with different $\beta$.
For example, for $N=4$, the two cases correspond to $a^{-1}\beta=1$ and $10$, for $N=6$, the two cases correspond to $a^{-1}\beta=1.5$ and $15$, etc.
Apart from that, $m=50\;{\rm MeV}$ is fixed and $am$ is different for different lattice extents.
The average of chiral condensation and the average of topological charge at $T=100\;{\rm MeV}$ and $10\;{\rm MeV}$ are shown in Figs.~\ref{fig:de-beta1} and \ref{fig:de-beta10}, respectively.

Compared with the finite volume effect, the chiral condensation convergent faster at lower temperatures.
Since the finite volume effect can be viewed as inferred cut off, while the discretization error can be viewed as the ultra-violate cut off, this result makes intuitive sense.
Furthermore, as the lattice spacing decreases, the convergence behavior of observables is not purely linear, exhibiting an additional exponential convergence or oscillatory convergence. 
Given that higher-order contributions become negligible for sufficiently small $a$, we perform the extrapolation exclusively on the $N=8,10,12$ data points while retaining the linear extrapolation.
The results of the linear extrapolations are also shown in Figs.~\ref{fig:de-beta1} and \ref{fig:de-beta10}.
The results indicate that linear extrapolation yields reasonable outcomes in most cases, with the exception of scenarios dominated by oscillatory or exponential convergence, where it overestimates discretization errors.
It can be expected that the extrapolation results become more reliable as long as a few number of additional data points are incorporated.

\subsection{\label{sec:5.2}The systematic errors of the method}

When using QMETTS, the system errors arise from approximating a thermal state with $|\phi _i\rangle/\sqrt{C_i}$, omitting the contribution of the $A$ matrix with coefficients $|c_j|<0.001$, and the Trotter expansion.
These deviations can be analyzed numerically by directly comparing the results of quantum simulation with exact diagonalization.
In the following analysis, statistical errors are also considered.
For both chiral condensation and fermion number, the statistical error of measuring the observables by repeating the circuit by $r$ times can be estimated as,
\begin{equation}
\begin{split}
&\epsilon _{st.}=\frac{1}{2}\frac{\sqrt{\sum _{i=1}^{16}\sum _{n=0}^{3}\left(\frac{C_{\phi _i}}{\sum _{j=1}^{16}C_{\phi _j} }\right)^2\left(1-\left\langle \phi_i \left(\frac{\beta}{2}\right)\right| \sigma _z(n) \left|\phi_i \left(\frac{\beta}{2}\right)\right\rangle ^2\right)}}{\sqrt{r}}.
\end{split}
\label{eq.5.1}
\end{equation}

\begin{figure}[htbp]
\includegraphics[width=0.48\hsize]{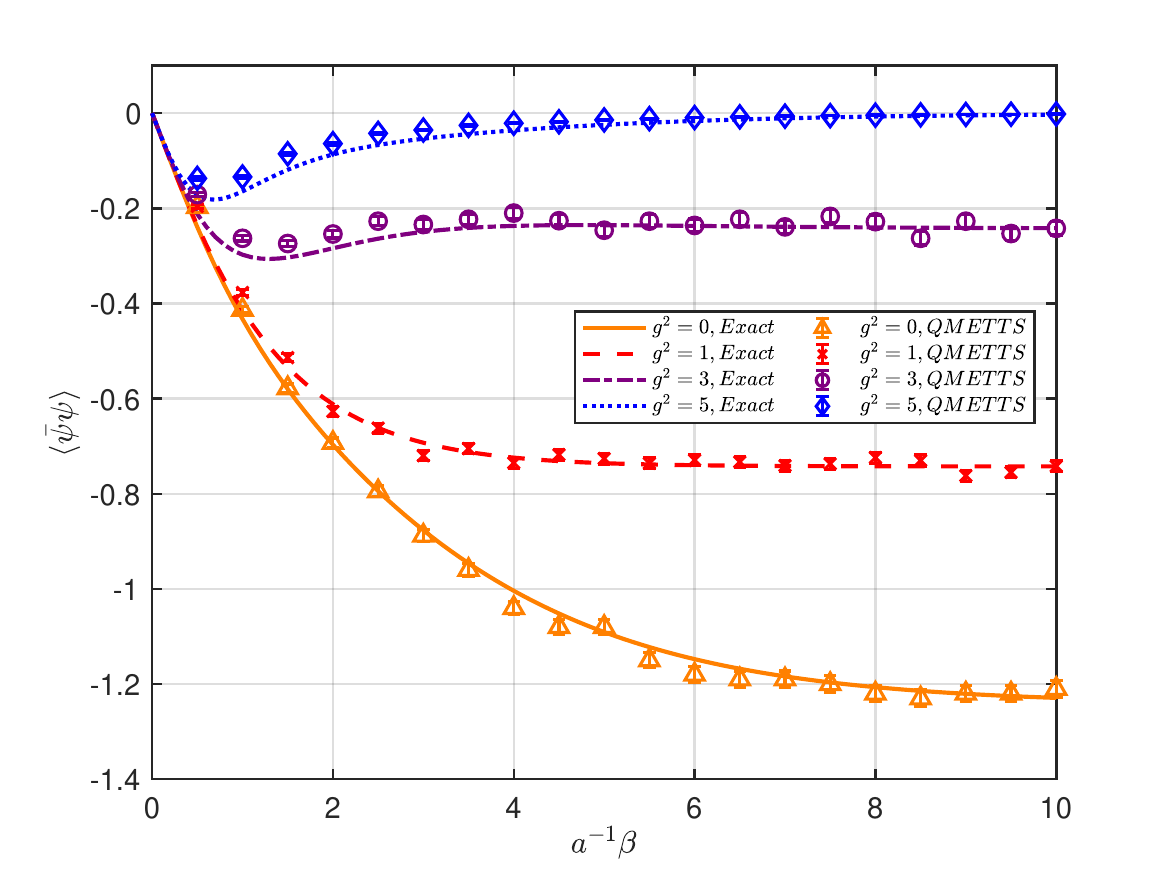}
\includegraphics[width=0.48\hsize]{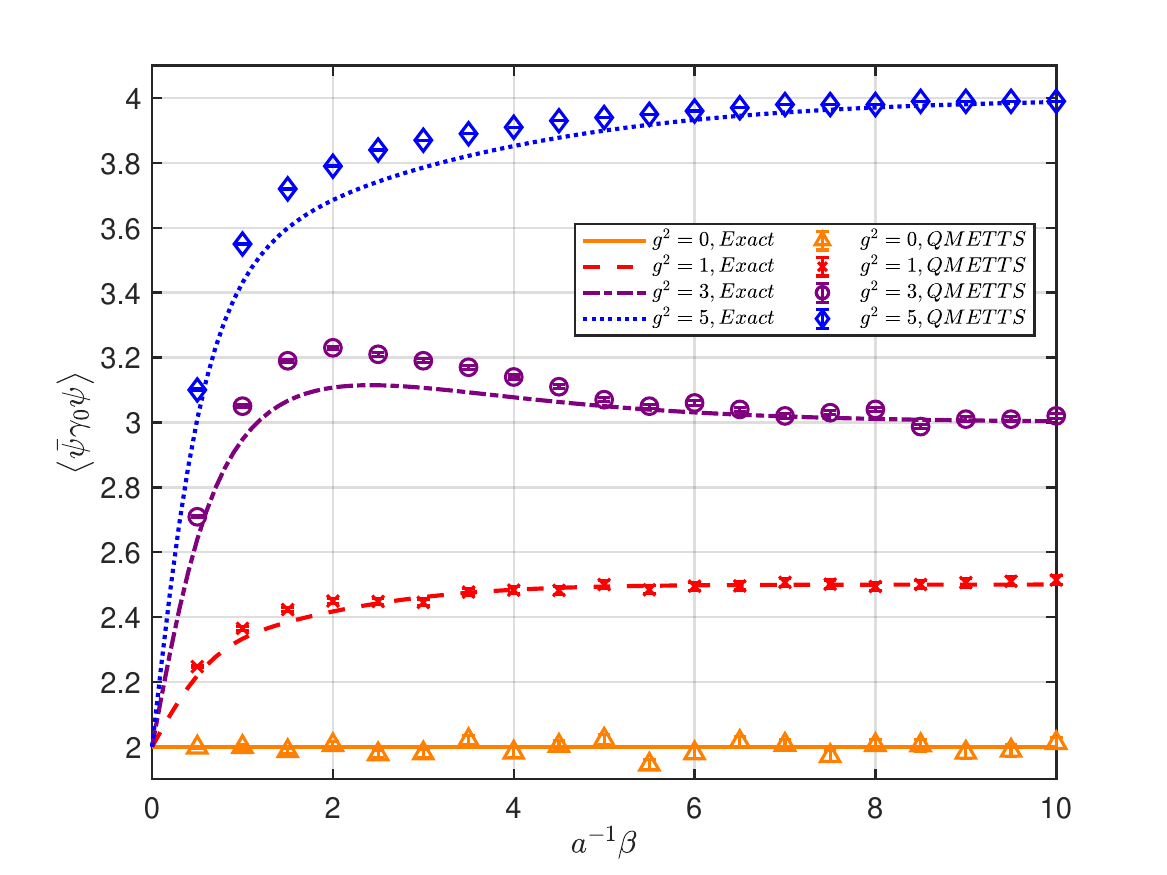}\\
\includegraphics[width=0.48\hsize]{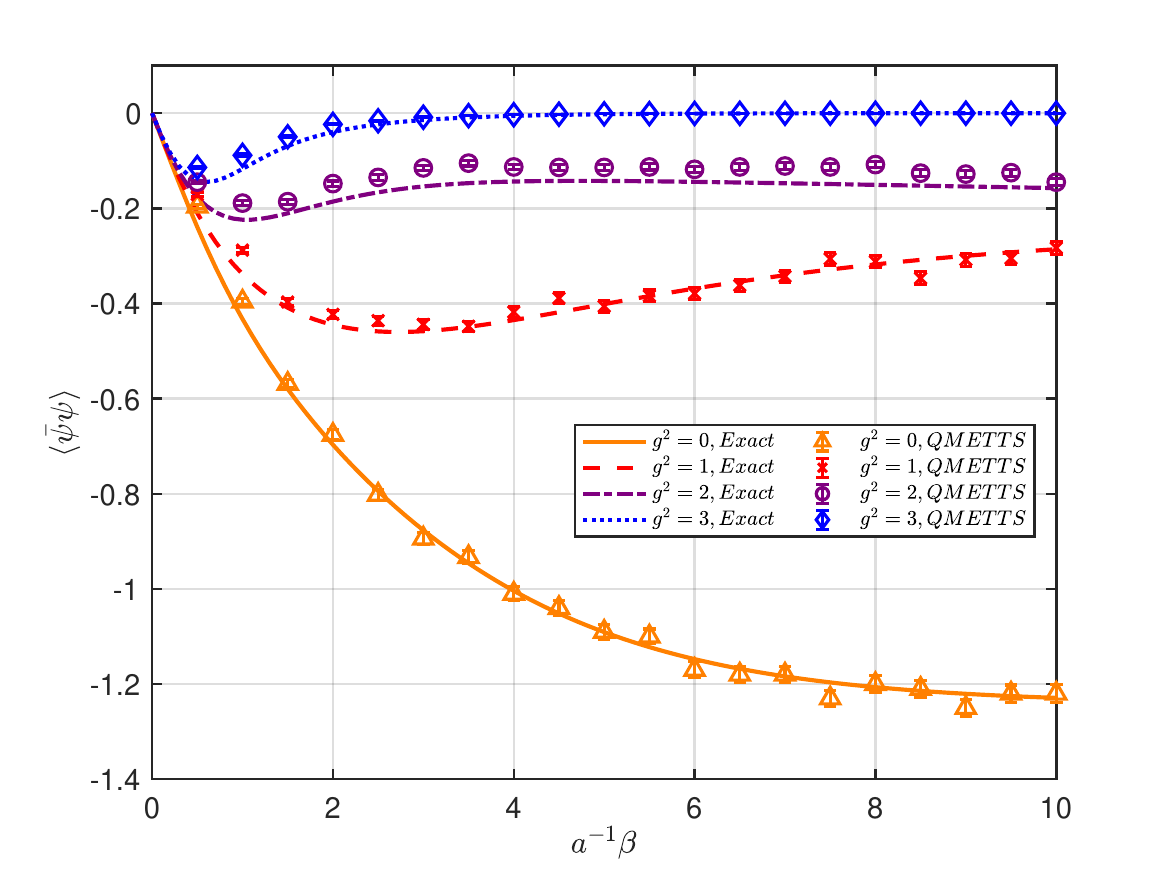}
\includegraphics[width=0.48\hsize]{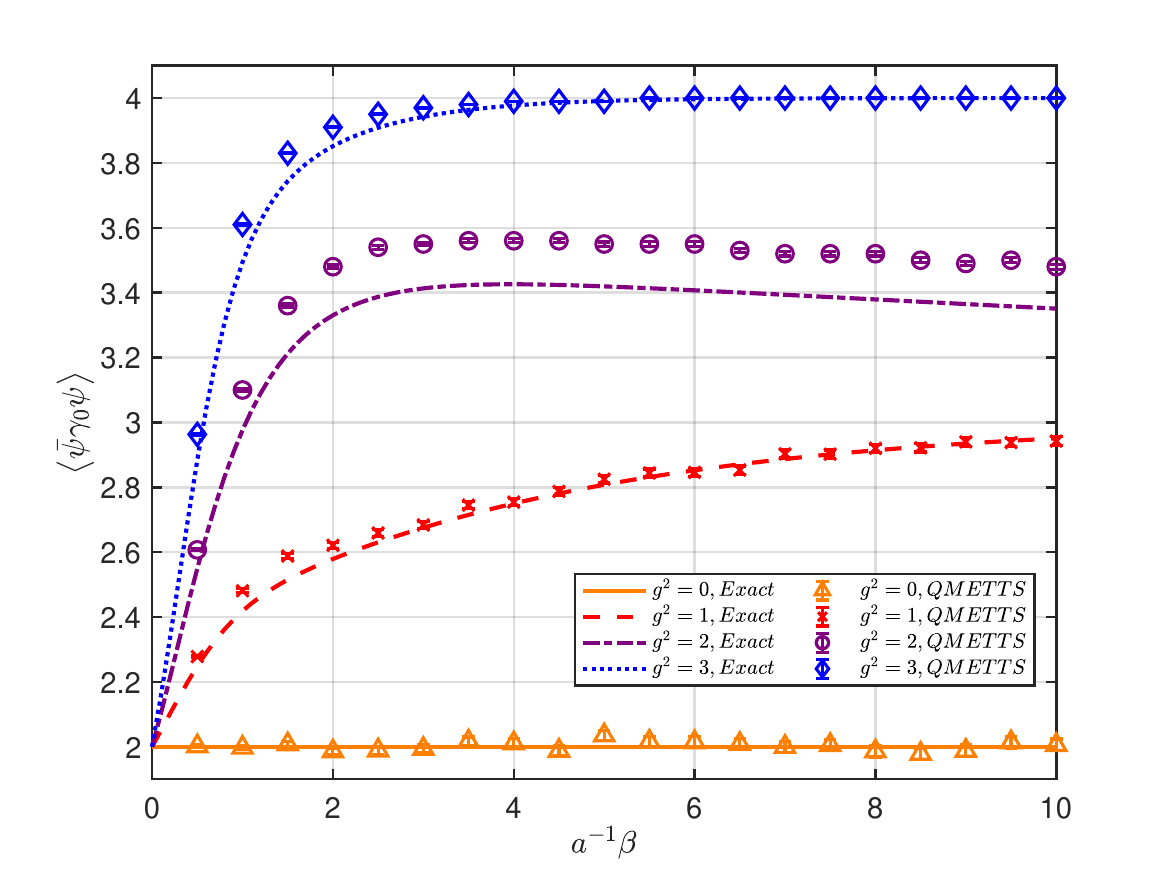}
\caption{\label{fig:differentbeta}.Comparison of QMETTS-derived observables with exact diagonalization results across different $g^2$ values as $\beta$ increases. 
Error bars represent statistical uncertainties from measurements estimated using Eq.~(\ref{eq.5.1}).
The upper panels are the results of $H_M$, the bottom panels are results of $H_E$, respectively.
With noise ignored, the deviation between QMETTS and exact diagonalization does not accumulate with the evolution of $\beta$.}
\end{figure}
For different $g^2$, the results of the observables measured with statistic errors by using QMETTS as well as the results from exact diagonalization along the evolution of $\beta$ are shown in Fig.~\ref{fig:differentbeta}.
It should be emphasized that the circuit is simulated on a classical computer, and the effect of noise is not considered.
It can be shown that in this case, the deviation between QMETTS and exact diagonalization does not accumulate with growing $\beta$.
This can be understood by one of the inherent advantages of QITE-based algorithms, that process of imaginary time evolution is akin to ``cooling'' a quantum system to its lowest energy configuration such that an initial state with any non-negligible overlap with the true ground state is guaranteed to evolve towards that ground state~\cite{Kolotouros:2024inn}.
This provides an advantage over other quantum-classical hybrid algorithms using classical optimizers, which can become trapped in local minima during optimization.

\begin{figure}[htbp]
\includegraphics[width=0.48\hsize]{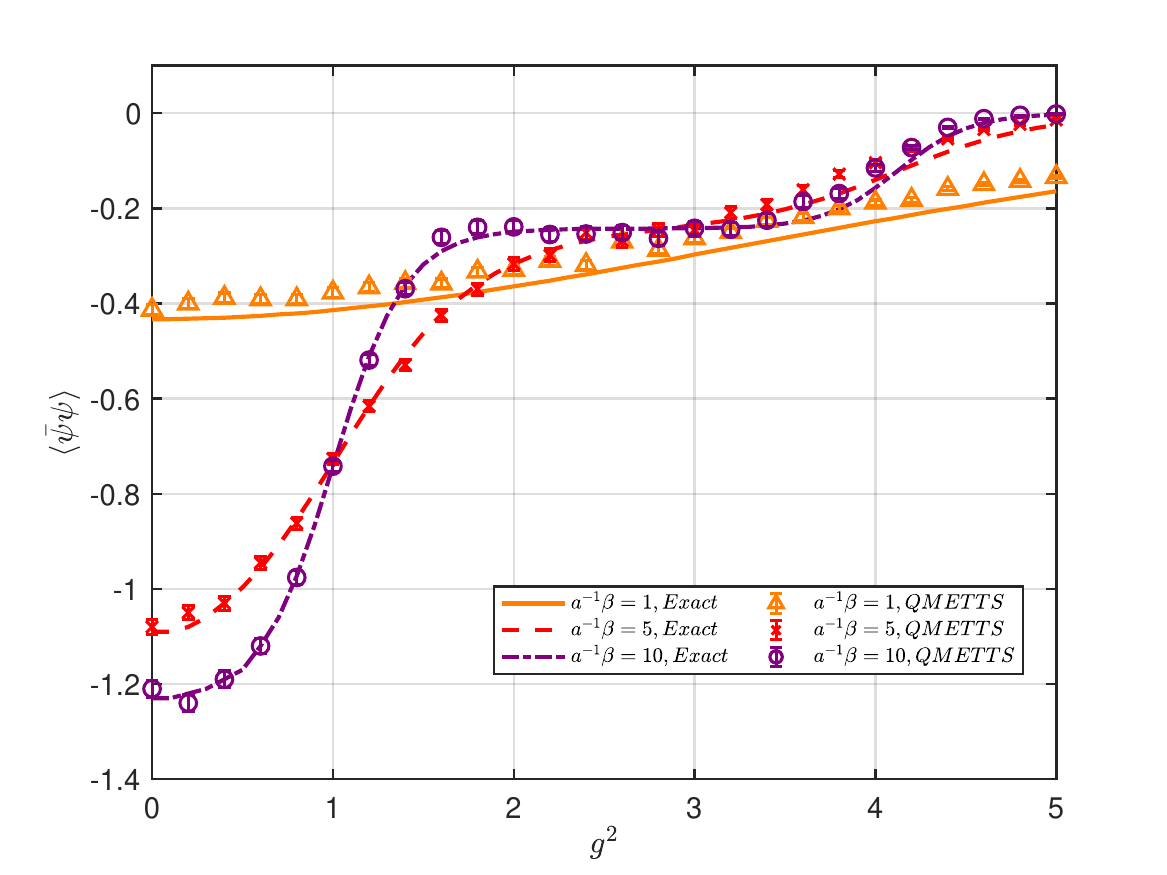}
\includegraphics[width=0.48\hsize]{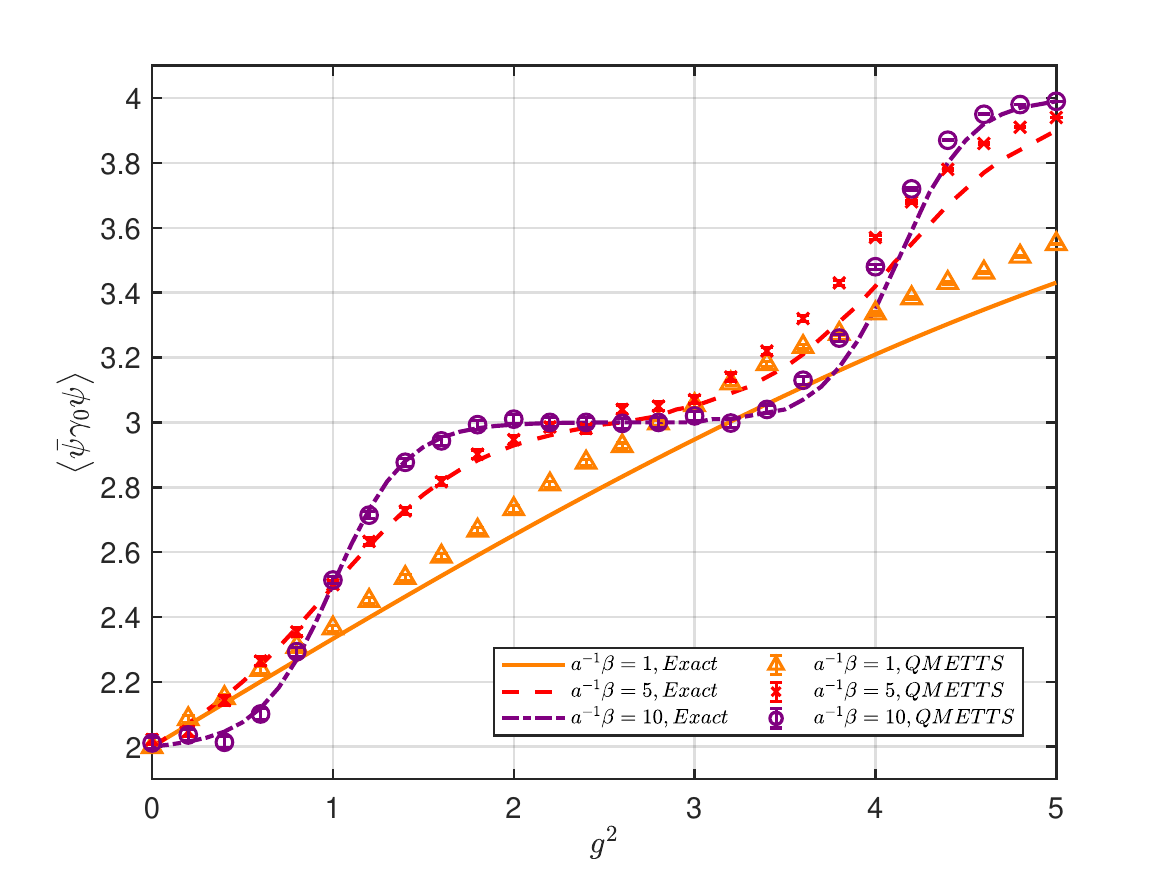}\\
\includegraphics[width=0.48\hsize]{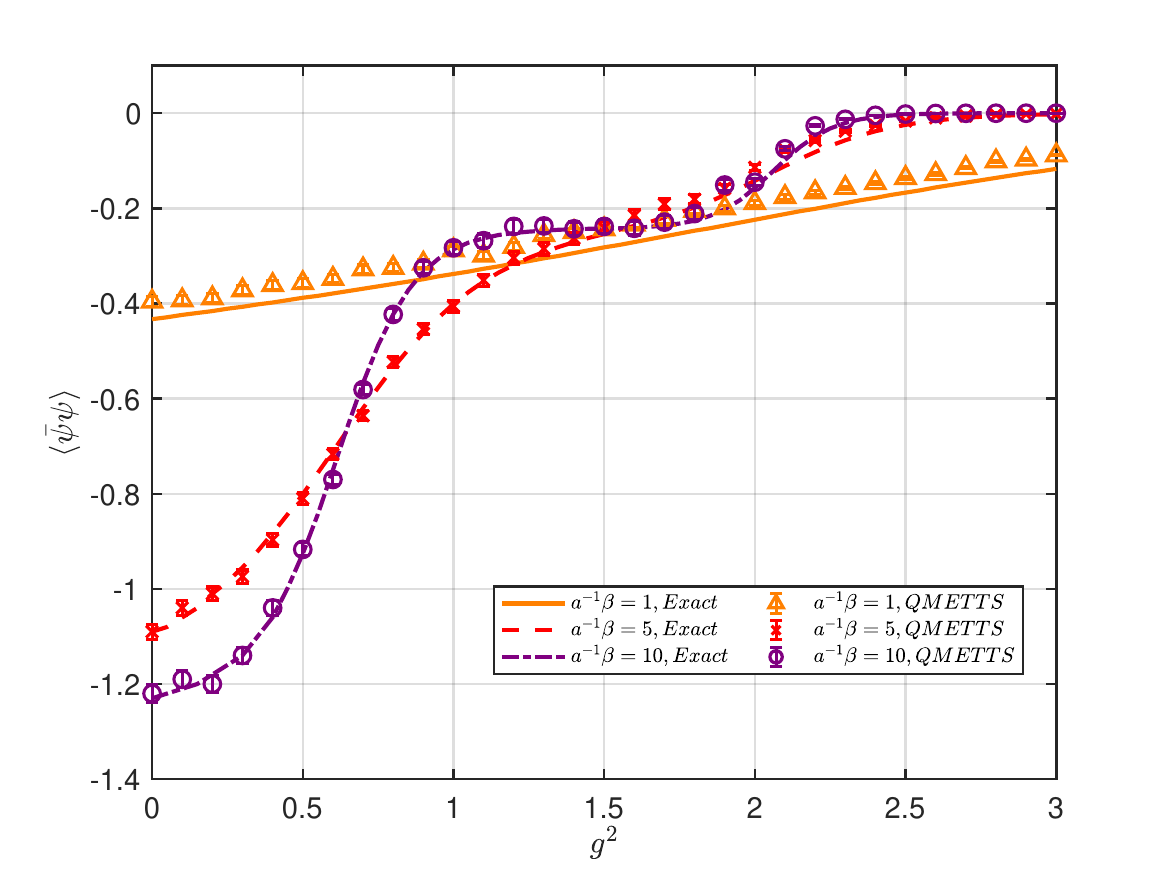}
\includegraphics[width=0.48\hsize]{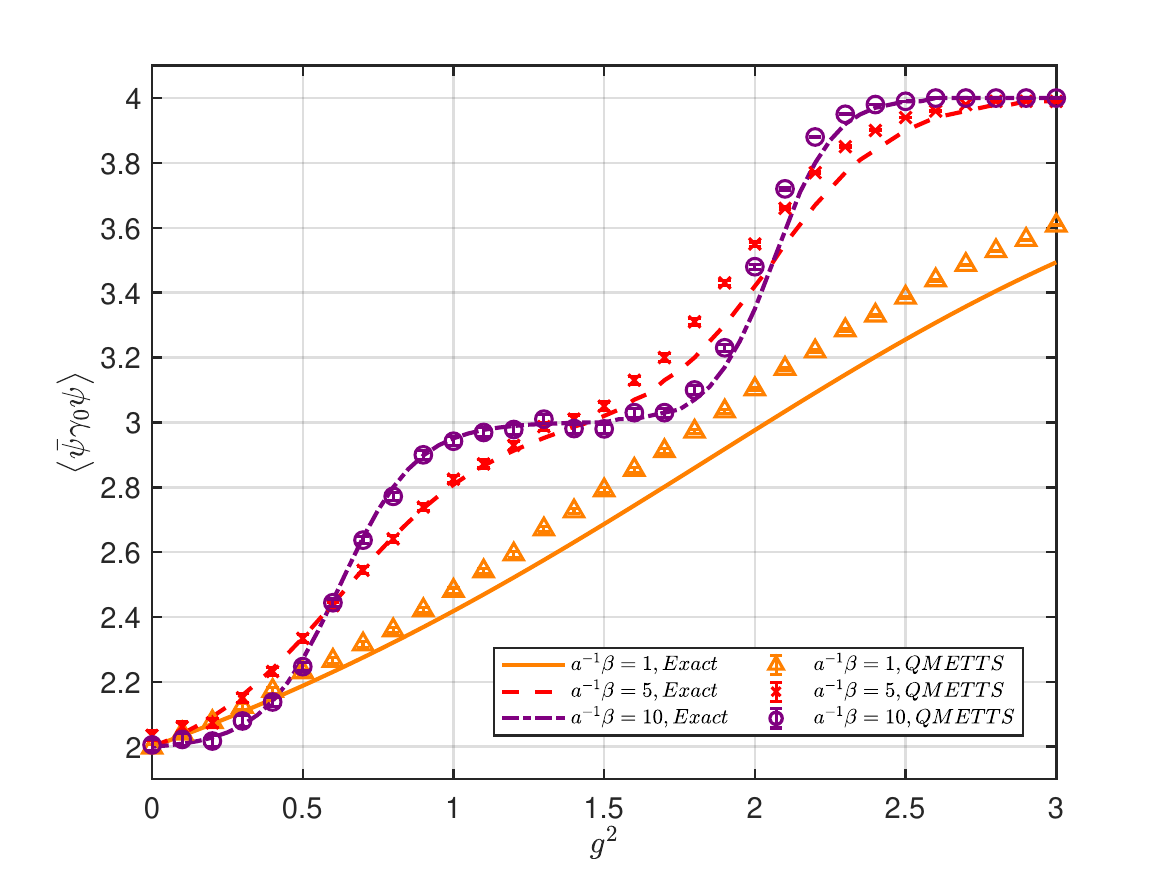}
\caption{\label{fig:differentcoupling}Same as Fig.~\ref{fig:differentbeta} but for evaluating to a same $\beta$ with growing coupling constant $g^2$.
It is found that, the deviation does not accumulate with growing $g^2$ either.
Note that, at $a^{-1}\beta=10$~($T=10\;{\rm MeV}$), the results are consistent with the exact diagonalization, indicating that the QMETTS can reproduce the behavior of massive Thirring model at low temperatures.}
\end{figure}
Observables computed via QMETTS and exact diagonalization are also compared across different $g^2$, as shown in Fig.~\ref{fig:differentcoupling}.
It can be seen that, the deviation does not accumulate with growing $g^2$ either.
The reason will be explored latter.
Within the range of coupling strengths $g^2$ considered, QITE proves applicable as a non-perturbative method.
The results at $a^{-1}\beta=10$~($T=10\;{\rm MeV}$) are consistent with the exact diagonalization, demonstrating the capabilities of QMETTS to capture the low temperature physics.

\begin{figure}[htbp]
\includegraphics[width=0.48\hsize]{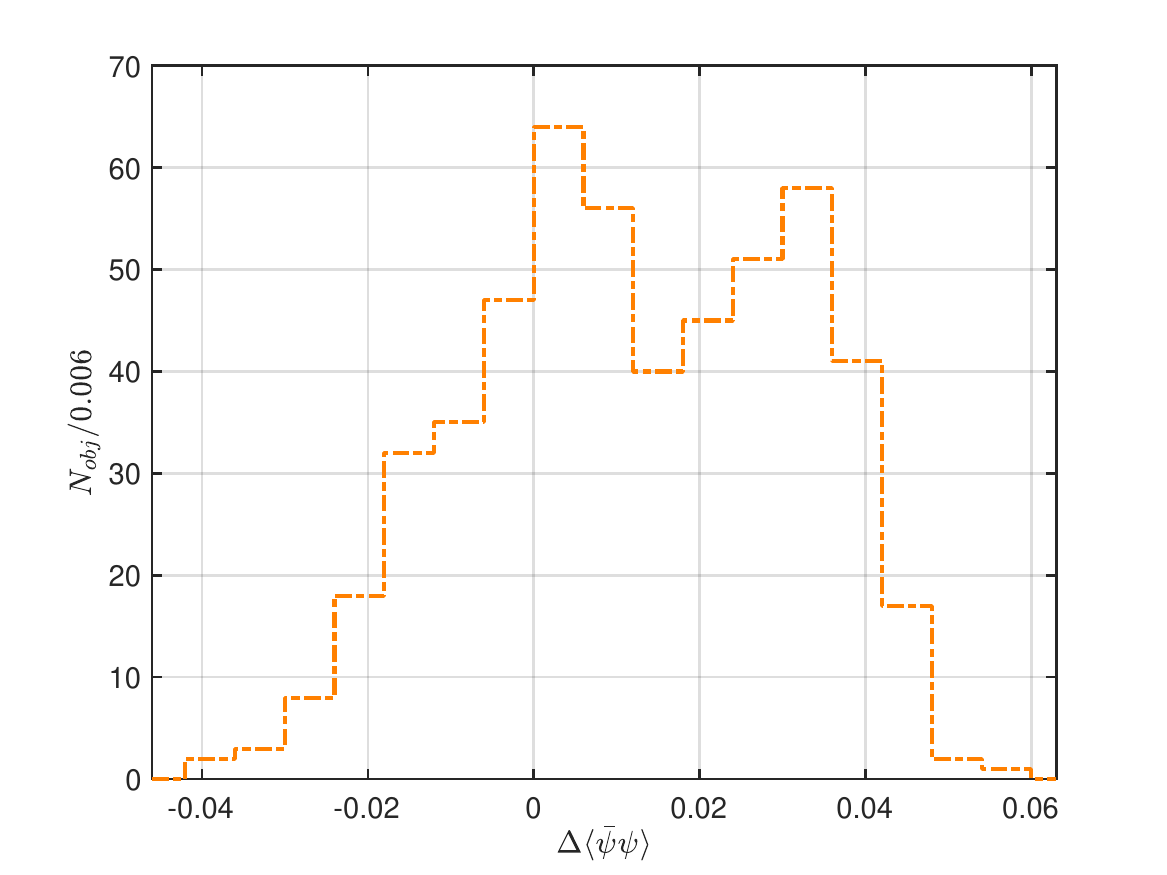}
\includegraphics[width=0.48\hsize]{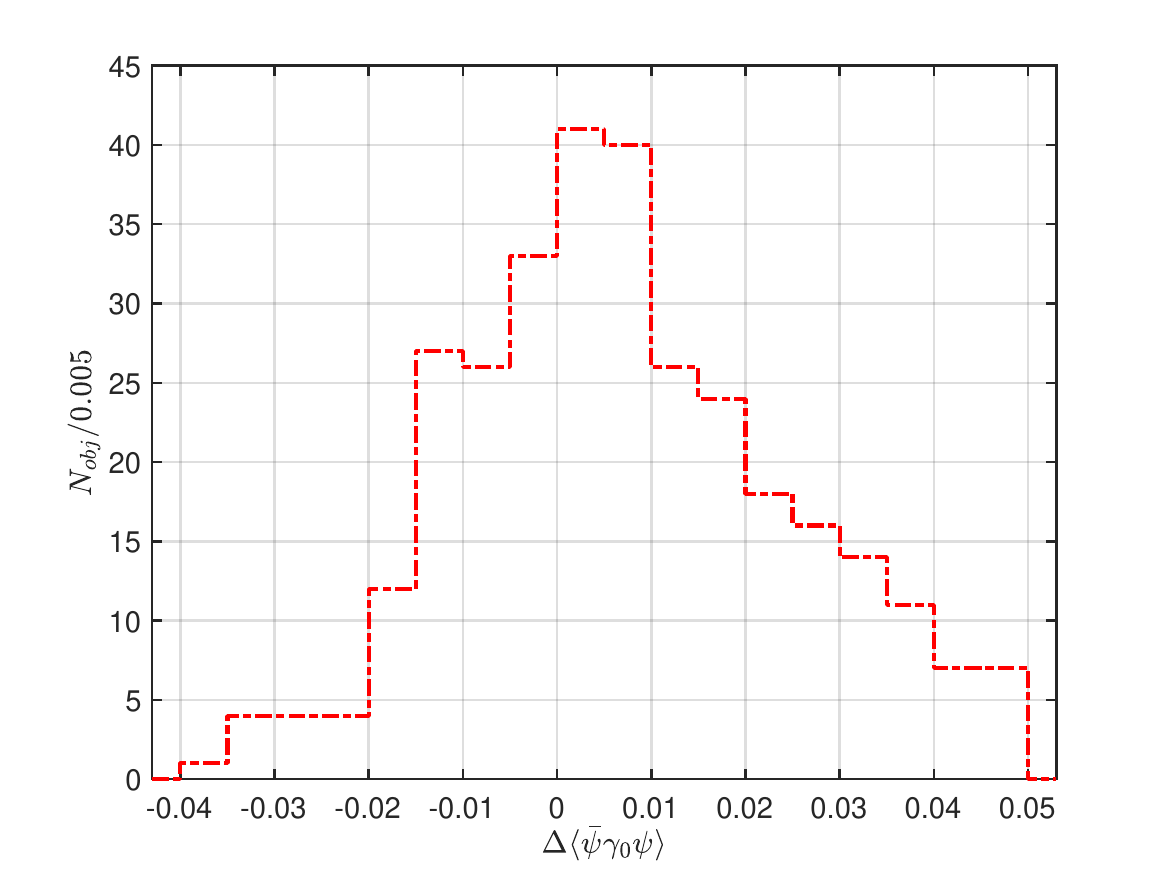}\\
\includegraphics[width=0.48\hsize]{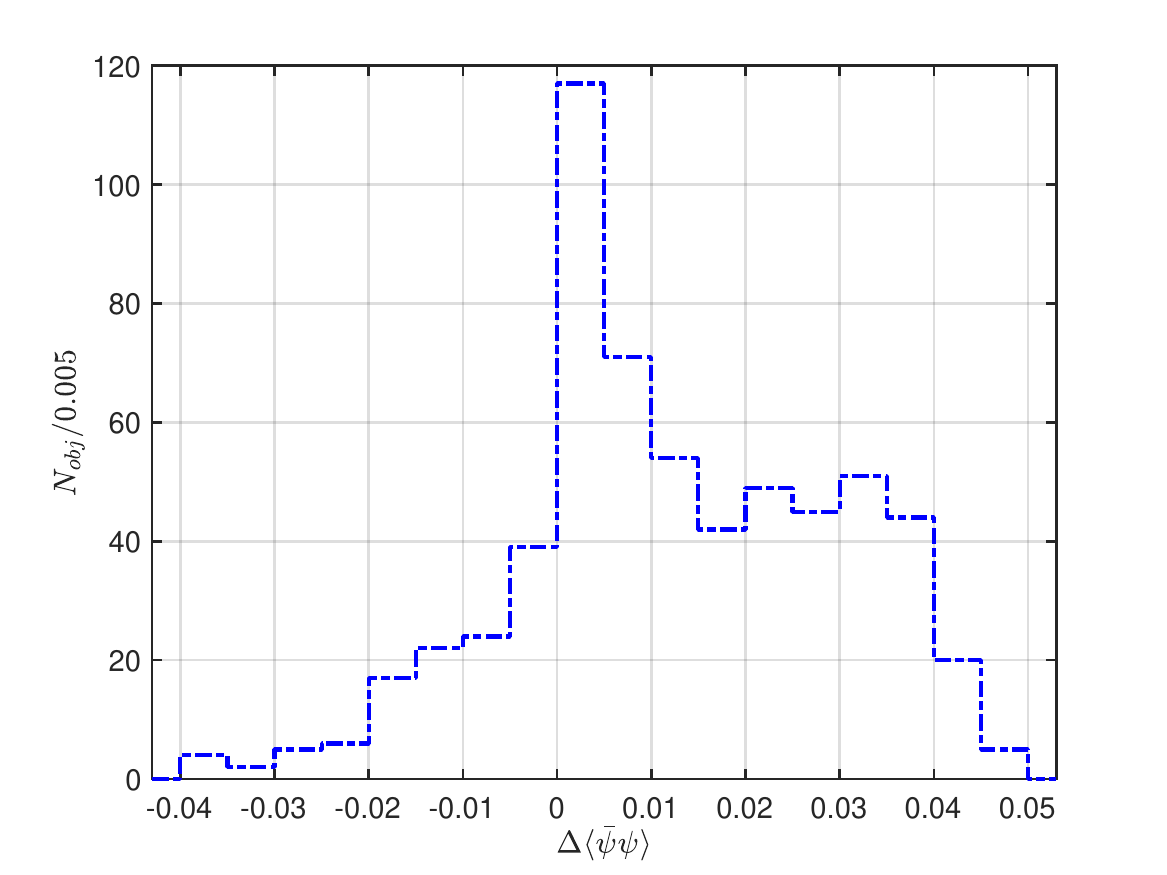}
\includegraphics[width=0.48\hsize]{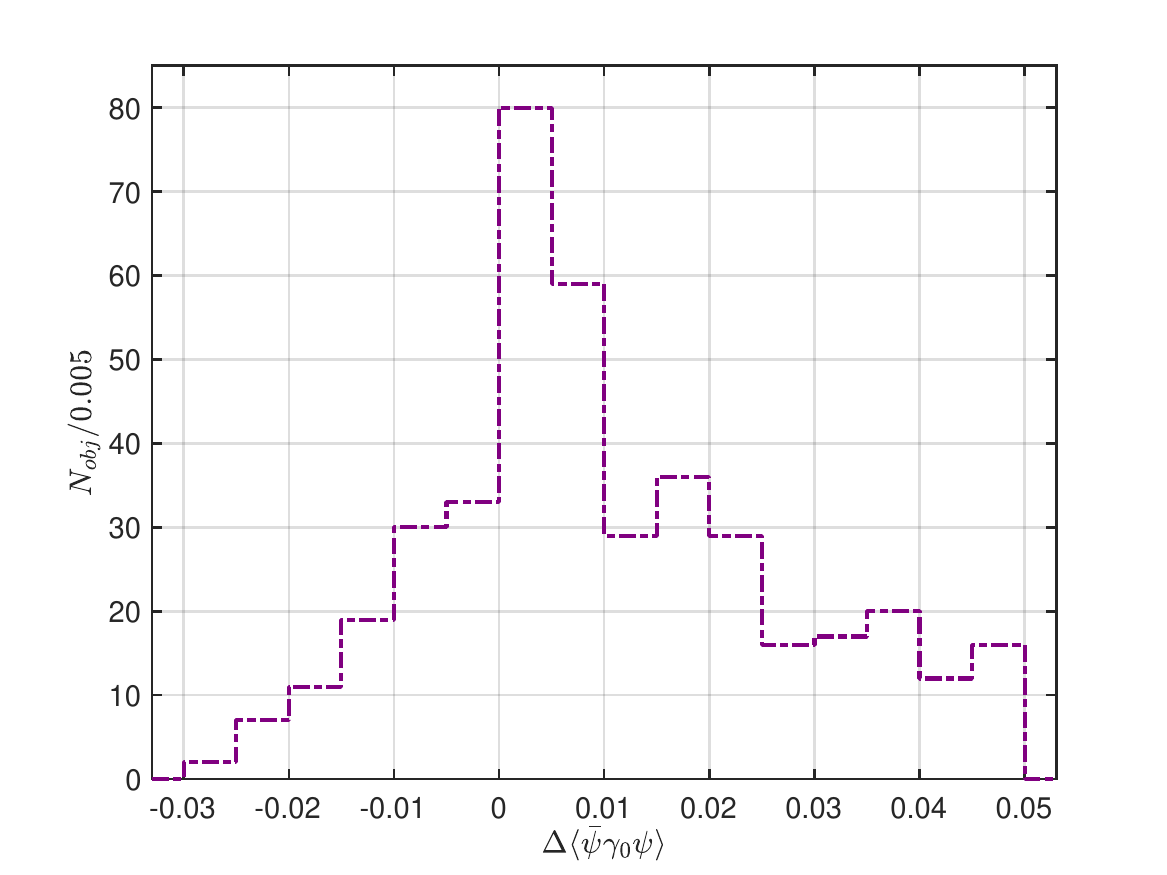}
\caption{\label{fig:errordistribution}Distributions of $\Delta \langle O \rangle$ for all measurements.
The upper panels are the results of $H_M$, the bottom panels are results of $H_E$, respectively.
The deviations distributed within the range $\left |\Delta \langle O \rangle \right| < 0.06$, and are approximately centered at zero.}
\end{figure}
In the case of $H_M$, with $K=20$ and $g^2=0,0.2,\ldots, 5.0$, there are $20\times 26$ data points measured for each observable, similarly, in the case of $H_E$, $g^2=0,0.1,\ldots, 3.0$, there are $20\times 31$ data points.
Defining $\Delta \langle O \rangle = \langle O \rangle _{\rm QMETTS}- \langle O \rangle _{\rm exact}$ where $\langle O \rangle _{\rm QMETTS}$ and $\langle O \rangle _{\rm exact}$ are values of observables obtained by using QMETTS and exact diagonalization, respectively, the distributions of $\Delta \langle O \rangle$ over the above mentioned data points are shown in Fig.~\ref{fig:errordistribution}.
It can be seen that, the deviations distributed within the range $\left |\Delta \langle O \rangle \right| < 0.06$, and are approximately centered at zero.
The characteristic pattern of error distribution validates the reliability of QMETTS.

\begin{figure}[htbp]
\begin{center}
\includegraphics[width=0.55\hsize]{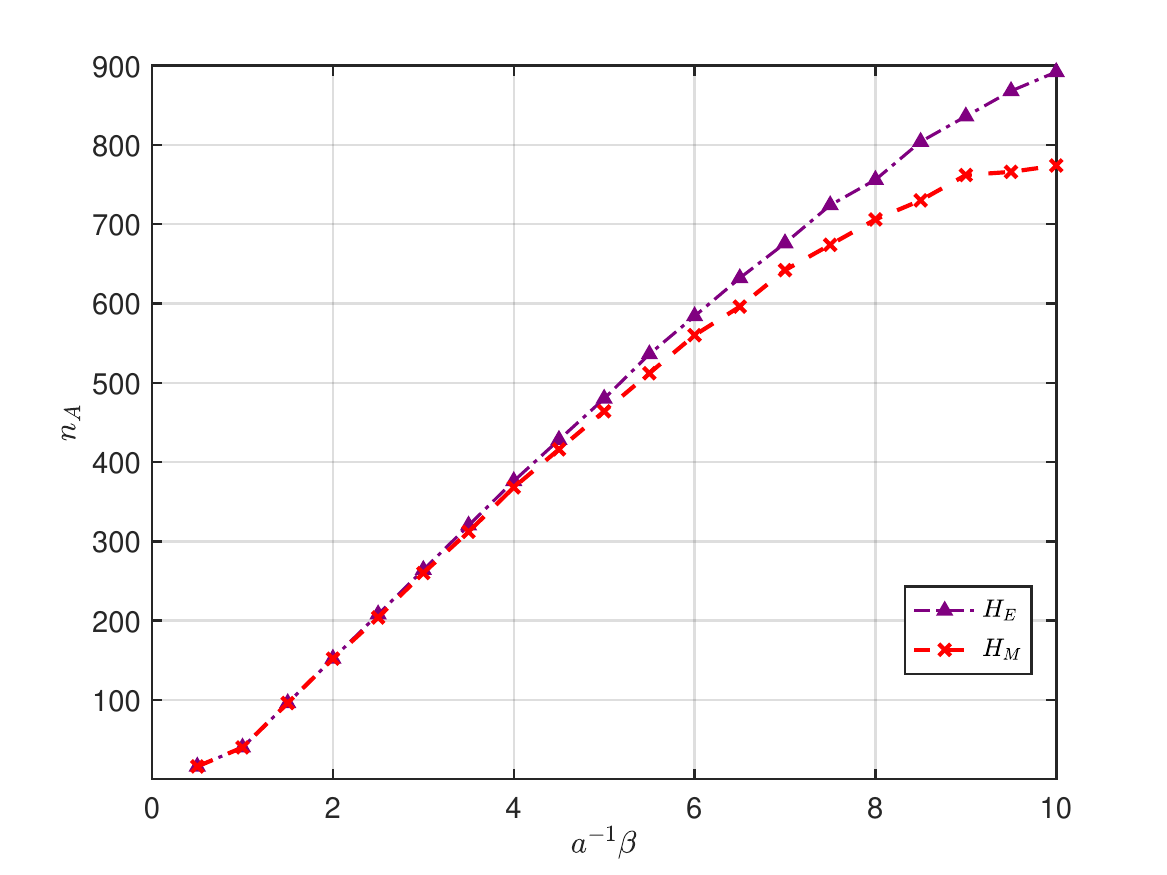}
\caption{\label{fig:na}The number of required Pauli tensor products during $\beta$ evolution. 
Given the variation in Pauli tensor counts depending on both $g^2$ and initial states, only the scenarios with the maximum required count are shown. 
The circuit depths remain substantial when evolving to larger $\beta$ values.}
\end{center}
\end{figure}
Another source of systematic error and computational cost not addressed is the calculation of the coefficients $c_j$.
This part of computational cost increases with the size of the system.
Since the operator $A$ is a linear combination of tensor products of Pauli matrices, the circuit depth can be roughly estimated by their number~(note that actual depth varies for different tensor products and depends on physical qubit topology, these variations are neglected). 
For different initial states $|\phi_i\rangle$ and different $g^2$, the required number of Pauli tensor products differs. 
We consider the maximum number of Pauli tensor products required when evolving to $a^{-1}\beta=10$ across all initial states, which is denoted as $n_A$.
$n_A$ serves as a rough estimate of both circuit depth and measurement complexity.
For $H_M$ and $H_E$, the $n_A$ throughout $\beta$ evolution are depicted in Fig.~\ref{fig:na}.
It can be seen that the circuit depths required for evolution to larger $\beta$ values remain beyond the capabilities of current quantum hardware.

Since the deviation between QMETTS and exact diagonalization arises from thermal state approximations and Trotter errors, the deviation can be estimated to correlate with $n_A$.
We find that to evolute to $a^{-1}\beta=10$, the values of $n_A$ remain comparable across different $g^2$. 
For $g^2=0$, $n_A=764$ are required. 
In the case of $H_M$ at $g^2=5$, the requirement is $n_A=774$. 
In the case of $H_E$ at $g^2=3$, up to $n_A=892$ terms are needed. 
Consequently, this explains why the deviation does not accumulate with growing $g^2$ as shown in Fig.~\ref{fig:differentcoupling}.

If we drop the periodic boundary conditions, for small $\beta$'s, the calculations of the coefficients can be simplified. 
Because at this point, the $H_M$ becomes an $XX$ model with an external transverse field in $1+1$ dimension, and $H_E$ becomes an $XXZ$ model with an external longitudinal field in $1+1$ dimension.
Each term in the Hamiltonian acts on at most 2 qubits (2-local), the $A$ operator only affects the domain around these two qubits been acted by the term~\cite{Motta:2019yya}. 
However this simplification disappears with the long-range correlation of the wave function with the increase of $\beta$. 
In this case, one can truncate the $A$ matrix using schemes such as inexact-QITE~\cite{Motta:2019yya} or extended local approximation or non-local approximation~\cite{Nishi:2021xhf}.
The truncation of the $A$ operator brings not only a simplification of the calculation of coefficients, but also a reduction of the circuit depth, as a result the computational complexity grows with the size of the correlation domain~(or the order of tensor products in $A$ operator in the case of inexact-QITE) instead of the size of the physics system to be simulated~\cite{Motta:2019yya,Nishi:2021xhf}.
Given our exploration into the low-temperature regime, where circuit depth becomes substantial, in our case implementation remains infeasible on real NISQ devices.
To address the issue of excessive circuit depth, several algorithmic approaches have been developed, for example by reincorporating variational machinery into the QITE algorithm, as exemplified by Refs.~\cite{Kumar:2024kij,Gomes:2021ckn}.
Improvements to QITE-algorithms are still active areas~\cite{Silva:2021jaj,Hejazi:2023fiq,Gluza:2024lqq}, and the massive thirring model can provide a test ground on those improvements.

\section{\label{sec:6}Summary}

In this work, we use the QMETTS approach to study the thermal properties of a massive Thirring model, which is a solvable system describing fermion interactions.
The key observables are the chiral condensate, which is associated with the chiral phase transition, and the fermion number, which corresponds to the topological charge in the dual sine-Gordon model.

The numerical results are obtained by using \verb"qiskit" to simulate the circuit on a classical computer, and with key components such as the operator coefficients and normalization factors computed exactly rather than through quantum measurements.
The results of the chiral condensation demonstrate that the chiral symmetry is broken at high temperatures, or at low temperatures and smaller coupling constants for the massive Thirring model. 
At low temperatures and large coupling constants, the chiral symmetry is restored. 
These results confirm the existence of a chiral phase transition in the massive Thirring model.
Furthermore, at low temperatures, the fermion number assumes integer values and exhibits quantized variation with the interaction strength.
This suggests that the ground state of the corresponding sine-Gordon model is situated within a distinct topological sector, indicating a topological phase transition. 
Additionally, a quantized alteration in the chiral condensation along with the fermion number is also observed at low temperatures.

A comparison of the results obtained from the QMETTS and exact diagonalization demonstrates that the utilization of QMETTS is capable of reproducing the outcomes of the exact diagonalization. 
In particular, with regard to the low-temperature case, it is evident that the QMETTS error do not accumulate with the evolution of $\beta$. 
This indicates that the QMETTS remains a valid tool for the study of low-temperature systems, as well as physical systems exhibiting topological phase transitions. 
In light of these findings, this work offers a valuable reference point for future QMETTS or QITE-based studies. 
As quantum computing technology advances, it is plausible that one will be able to surmount the sign problem in lattice QCD with the aid of quantum computing.

\acknowledgments

This work was supported in part by the National Natural Science Foundation of China under Grants No.~12147214, the Natural Science Foundation of the Liaoning Scientific Committee No.~LJKZ0978.

\appendix

\section{\label{sec:ap1}Details on the \texorpdfstring{$\mathcal{O}(a)$}{O(a)} terms}

The discretization of fermions into staggered fermions distributed over the two sites can result in the emergence of terms within a pair of sites. 
This leads to the presence of terms at every two sites. 
These spatially discontinuous terms can be reduced to spatially continuous terms, accompanied by an additional term of higher order in terms of the lattice spacing $a$. 
In this work, we choose to ignore these higher-order terms in order to maintain the spatial continuity of the Hamiltonian.
With,
\begin{equation}
\begin{split}
&\partial _h \chi(n) = \frac{1}{a}\left(\chi(n+2)-\chi(n)\right),\\
\end{split}
\end{equation}
and Eqs.~(\ref{eq.2.2}), (\ref{eq.2.3}) and (\ref{eq.2.5}), it can be verified,
\begin{equation}
\begin{split}
&\sum _h \bar{\psi}(h)i\gamma _1\partial _x \psi(h)
= -\frac{i}{a^2} \sum _n \left(\chi^{\dagger}(n)\chi (n+1) - \chi ^{\dagger}(n+1)\chi (n)\right),\\
&a\sum _h \bar{\psi}(h) \psi(h) = \sum _{n} (-1)^n \chi^{\dagger} (n)\chi (n),\\
&a\sum _h \bar{\psi}(h) \gamma _0 \psi(h) =\sum _{n} \chi^{\dagger} (n)\chi (n),\\
&a^2\sum _h (\bar{\psi}(h) \psi(h) )^2=
\sum _n \left\{(\chi ^{\dagger}(n)\chi (n))^2-(\chi ^{\dagger}(n)\chi (n)) (\chi ^{\dagger}(n+1)\chi (n+1))\right\},\\
&-a\sum _{even\;n}\left\{(\partial _h\chi^{\dagger}(n))\chi(n)\chi^{\dagger}(n+1)\chi(n+1)
+\chi^{\dagger}(n+2)(\partial _h\chi(n))\chi^{\dagger}(n+1)\chi(n+1)\right\},\\
&a^2\sum _h (\bar{\psi}(h) \gamma _0 \psi(h) )^2=
\sum _n \left((\chi ^{\dagger}(n)\chi (n))^2+(\chi ^{\dagger}(n)\chi (n)) (\chi ^{\dagger}(n+1)\chi (n+1))\right)\\
&+a\sum _{even\;n}\left\{(\partial _h\chi^{\dagger}(n))\chi(n)\chi^{\dagger}(n+1)\chi(n+1)
+\chi^{\dagger}(n+2)(\partial _h\chi(n))\chi^{\dagger}(n+1)\chi(n+1)\right\},\\
&a^2\sum _h (\bar{\psi}(h) \gamma _1 \psi(h) )^2=
\frac{1}{2}\sum _n \left(\chi^{\dagger}(n)\chi (n+1)+\chi^{\dagger} (n+1)\chi (n)\right)^2\\
&-\frac{a}{2}\sum _{even\;n}\left((\partial _h \chi^{\dagger}(n))\chi (n+1)\chi ^{\dagger}(n)\chi (n+1)
 +\chi ^{\dagger}(n+2)\chi (n+1)(\partial _h\chi ^{\dagger}(n))\chi (n+1)\right.\\
&\left.+\chi^{\dagger}(n+1)(\partial _h \chi (n))\chi ^{\dagger}(n)\chi (n+1)
+\chi ^{\dagger}(n+1)\chi (n+2)\chi ^{\dagger}(n+1)(\partial _h \chi (n))\right)\\
&-\frac{a}{2}\sum _{even\;n}\left\{(\partial _h\chi^{\dagger}(n))\chi (n+1)\chi ^{\dagger}(n)\chi (n+1)
+\chi ^{\dagger}(n+1)\chi (n+2)\chi ^{\dagger}(n+1)(\partial _h \chi (n))\right.\\
&\left.+\chi^{\dagger}(n+1)(\partial _h \chi (n))\chi ^{\dagger}(n)\chi (n+1)
+\chi ^{\dagger}(n+1)\chi (n+2)(\partial _h \chi ^{\dagger}(n))\chi (n+1)\right\}.\\
\end{split}
\end{equation}
Neglecting the $\mathcal{O}(a)$ terms,
\begin{equation}
\begin{split}
a^2\sum _h (\bar{\psi}(h) \psi(h) )^2&=
\sum _n (\chi ^{\dagger}(n)\chi (n)\left(\chi ^{\dagger}(n)\chi (n)-\chi ^{\dagger}(n+1)\chi (n+1)\right)+\mathcal{O}(a),\\
a^2\sum _h (\bar{\psi}(h) \gamma _0 \psi(h) )^2&=
\sum _n (\chi ^{\dagger}(n)\chi (n)\left(\chi ^{\dagger}(n)\chi (n)+\chi ^{\dagger}(n+1)\chi (n+1)\right)+\mathcal{O}(a),\\
a^2\sum _h (\bar{\psi}(h) \gamma _1 \psi(h) )^2&=
\frac{1}{2}\sum _n \left(\chi^{\dagger}(n)\chi (n+1)+\chi^{\dagger} (n+1)\chi (n)\right)^2+\mathcal{O}(a).\\
\end{split}
\end{equation}

Then Eqs.~(\ref{eq.2.11}) and (\ref{eq.2.12}) can be obtained by substituting $\chi$ with Eq.~(\ref{eq.2.6}).

\section{\label{sec:ap2}Measurements to obtain \texorpdfstring{$S$}{S} and \texorpdfstring{$b$}{b}}

Since $|\phi(\beta)\rangle$ is a quantum state, the matrix elements of $S$ and $b$ cannot be obtained classically.
Therefore, the matrix elements of $S$ and $b$ should be determined through measurements.
Since the Hamiltonian can also be expanded using $aH=\sum _j h_j\hat{\sigma} _{j}$, essentially, one needs a quick construction of $\hat{\sigma} _{j_1}\hat{\sigma} _{j_2}+\left(\hat{\sigma} _{j_1}\hat{\sigma} _{j_2}\right)^T$ and $-{\rm i}\left(\hat{\sigma} _{j_1}\hat{\sigma} _{j_2}-\hat{\sigma} _{j_2}\hat{\sigma} _{j_1}\right)$.
Defining,
\begin{equation}
\begin{split}
M_a=\begin{pmatrix}0 & 1 & 2 & 3 \\ 1 & 0 & 3 & 2 \\ 2 & 3 & 0 & 1\\3 & 2 & 1 & 0\end{pmatrix},\;\;&M_b=\begin{pmatrix}0 & 0 & 0 & 0 \\ 0 & 0 & 1 & 1 \\ 0 & 1 & 0 & 1\\ 0 & 1 & 1 & 0\end{pmatrix},\\
M_c=\begin{pmatrix}0 & 0 & 0 & 0 \\ 0 & 0 & 1 & 0 \\ 0 & 0 & 0 & 1\\ 0 & 1 & 0 & 0\end{pmatrix},\;\;&M_d=\begin{pmatrix}0 & 0 & 1 & 0 \\ 0 & 0 & 0 & 1 \\ 1 & 0 & 0 & 0\\ 0 & 1 & 0 & 0\end{pmatrix}.\\
\end{split}
\label{eq.b.1}
\end{equation}
Denoting the separate indices as $x,y$ for $j_1$ and $j_2$, respectively, i.e., $j_1=(x_1,x_2,\ldots)$ and $j_2=(y_1,y_2,\ldots)$, it can be verified,
\begin{equation}
\begin{split}
&\hat{\sigma} _{j_1}\hat{\sigma} _{j_2}+\left(\hat{\sigma} _{j_1}\hat{\sigma} _{j_2}\right)^T=(-1)^{\sum _{i_1} \left(M_c^T\right)^{x_{i_1},y_{i_1}}}{\rm i}^{\sum _{i_2} M_b^{x_{i_2},y_{i_2}}}
 \left(1+(-1)^{\sum _{i_3} M_d^{x_{i_3},y_{i_3}}}\right)\prod _{l}\sigma^{M_a^{x_l,y_l}} (l),
\end{split}
\label{eq.b.2}
\end{equation}
and,
\begin{equation}
\begin{split}
&-{\rm i}\left(\hat{\sigma} _{j_1}\hat{\sigma} _{j_2}-\hat{\sigma} _{j_2}\hat{\sigma} _{j_1}\right)=(-1)^{\sum _{i_1} M_c^{x_{i_1},y_{i_1}}}{\rm i}^{\left(1+\sum _{i_2} M_b^{x_{i_2},y_{i_2}}\right)}
 \left(1-(-1)^{\sum _{i_3} M_b^{x_{i_3},y_{i_3}}}\right)\prod _{l}\sigma^{M_a^{x_l,y_l}} (l),
\end{split}
\label{eq.b.3}
\end{equation}
where $M^{x,y}$ is the matrix element of $M$ at index $(x,y)$.
$S+S^T$ in Eq.~(\ref{eq.3.4}) can be obtained directly using Eq.~(\ref{eq.b.2}).
For the components of $b$ defined in Eq.~(\ref{eq.3.5}), \raggedright 
$b_j= \sum _{j'}h_{j'}\left\langle \phi|-{\rm i}\left(\hat{\sigma} _{j}\hat{\sigma} _{j'}-\hat{\sigma} _{j'}\hat{\sigma} _{j}\right)|\phi\right\rangle/\sqrt{C(\Delta \beta)}$, 
\normalfont
where $h_{j'}$ are the dimensionless coefficients in the expansion of Hamiltonian.

\bibliography{thirring}

\providecommand{\href}[2]{#2}\begingroup\raggedright\begin{thebibliography}{10}

\bibitem{Arute:2019zxq}
F.~Arute et~al., \emph{{Quantum supremacy using a programmable superconducting
  processor}}, \href{https://doi.org/10.1038/s41586-019-1666-5}{\emph{Nature}
  {\bfseries 574} (2019) 505}
  [\href{https://arxiv.org/abs/1910.11333}{{\ttfamily 1910.11333}}].

\bibitem{Preskill:2018jim}
J.~Preskill, \emph{{Quantum Computing in the NISQ era and beyond}},
  \href{https://doi.org/10.22331/q-2018-08-06-79}{\emph{Quantum} {\bfseries 2}
  (2018) 79} [\href{https://arxiv.org/abs/1801.00862}{{\ttfamily 1801.00862}}].

\bibitem{Chen:2021num}
Z.~Chen, K.~J. Satzinger, J.~Atalaya, A.~N. Korotkov, A.~Dunsworth, D.~Sank
  et~al., \emph{Exponential suppression of bit or phase errors with cyclic
  error correction},
  \href{https://doi.org/10.1038/s41586-021-03588-y}{\emph{Nature} {\bfseries
  595} (2021) 383} [\href{https://arxiv.org/abs/2102.06132}{{\ttfamily
  2102.06132}}].

\bibitem{Fang:2024ple}
Y.~Fang, C.~Gao, Y.-Y. Li, J.~Shu, Y.~Wu, H.~Xing et~al., \emph{{Quantum
  Frontiers in High Energy Physics}},  11, 2024.

\bibitem{Scott:2024txs}
J.~L. Scott, Z.~Dong, T.~Kim, K.~Kong and M.~Park, \emph{{Hybrid
  quantum-classical approach for combinatorial problems at hadron colliders}},
  10, 2024, \href{https://arxiv.org/abs/2410.22417}{{\ttfamily 2410.22417}}.

\bibitem{Roggero:2018hrn}
A.~Roggero and J.~Carlson, \emph{{Dynamic linear response quantum algorithm}},
  \href{https://doi.org/10.1103/PhysRevC.100.034610}{\emph{Phys. Rev. C}
  {\bfseries 100} (2019) 034610}
  [\href{https://arxiv.org/abs/1804.01505}{{\ttfamily 1804.01505}}].

\bibitem{Roggero:2019myu}
A.~Roggero, A.~C.~Y. Li, J.~Carlson, R.~Gupta and G.~N. Perdue, \emph{{Quantum
  Computing for Neutrino-Nucleus Scattering}},
  \href{https://doi.org/10.1103/PhysRevD.101.074038}{\emph{Phys. Rev. D}
  {\bfseries 101} (2020) 074038}
  [\href{https://arxiv.org/abs/1911.06368}{{\ttfamily 1911.06368}}].

\bibitem{Atas:2021ext}
Y.~Y. Atas, J.~Zhang, R.~Lewis, A.~Jahanpour, J.~F. Haase and C.~A. Muschik,
  \emph{{SU(2) hadrons on a quantum computer via a variational approach}},
  \href{https://doi.org/10.1038/s41467-021-26825-4}{\emph{Nature Commun.}
  {\bfseries 12} (2021) 6499}
  [\href{https://arxiv.org/abs/2102.08920}{{\ttfamily 2102.08920}}].

\bibitem{Lamm:2019uyc}
{\scshape NuQS} collaboration, \emph{{Parton physics on a quantum computer}},
  \href{https://doi.org/10.1103/PhysRevResearch.2.013272}{\emph{Phys. Rev.
  Res.} {\bfseries 2} (2020) 013272}
  [\href{https://arxiv.org/abs/1908.10439}{{\ttfamily 1908.10439}}].

\bibitem{Li:2021kcs}
{\scshape QuNu} collaboration, \emph{{Partonic collinear structure by quantum
  computing}}, \href{https://doi.org/10.1103/PhysRevD.105.L111502}{\emph{Phys.
  Rev. D} {\bfseries 105} (2022) L111502}
  [\href{https://arxiv.org/abs/2106.03865}{{\ttfamily 2106.03865}}].

\bibitem{Perez-Salinas:2020nem}
A.~P\'erez-Salinas, J.~Cruz-Martinez, A.~A. Alhajri and S.~Carrazza,
  \emph{{Determining the proton content with a quantum computer}},
  \href{https://doi.org/10.1103/PhysRevD.103.034027}{\emph{Phys. Rev. D}
  {\bfseries 103} (2021) 034027}
  [\href{https://arxiv.org/abs/2011.13934}{{\ttfamily 2011.13934}}].

\bibitem{Jordan:2011ci}
S.~P. Jordan, K.~S.~M. Lee and J.~Preskill, \emph{{Quantum Computation of
  Scattering in Scalar Quantum Field Theories}}, {\emph{Quant. Inf. Comput.}
  {\bfseries 14} (2014) 1014}
  [\href{https://arxiv.org/abs/1112.4833}{{\ttfamily 1112.4833}}].

\bibitem{Mueller:2019qqj}
N.~Mueller, A.~Tarasov and R.~Venugopalan, \emph{{Deeply inelastic scattering
  structure functions on a hybrid quantum computer}},
  \href{https://doi.org/10.1103/PhysRevD.102.016007}{\emph{Phys. Rev. D}
  {\bfseries 102} (2020) 016007}
  [\href{https://arxiv.org/abs/1908.07051}{{\ttfamily 1908.07051}}].

\bibitem{Guan:2020bdl}
W.~Guan, G.~Perdue, A.~Pesah, M.~Schuld, K.~Terashi, S.~Vallecorsa et~al.,
  \emph{{Quantum Machine Learning in High Energy Physics}},
  \href{https://doi.org/10.1088/2632-2153/abc17d}{\emph{Mach. Learn. Sci.
  Tech.} {\bfseries 2} (2021) 011003}
  [\href{https://arxiv.org/abs/2005.08582}{{\ttfamily 2005.08582}}].

\bibitem{Wu:2021xsj}
S.~L. Wu et~al., \emph{{Application of quantum machine learning using the
  quantum kernel algorithm on high energy physics analysis at the LHC}},
  \href{https://doi.org/10.1103/PhysRevResearch.3.033221}{\emph{Phys. Rev.
  Res.} {\bfseries 3} (2021) 033221}
  [\href{https://arxiv.org/abs/2104.05059}{{\ttfamily 2104.05059}}].

\bibitem{Zhang:2023ykh}
S.~Zhang, Y.-C. Guo and J.-C. Yang, \emph{{Optimize the event selection
  strategy to study the anomalous quartic gauge couplings at muon colliders
  using the support vector machine and quantum support vector machine}},
  \href{https://doi.org/10.1140/epjc/s10052-024-13208-4}{\emph{Eur. Phys. J. C}
  {\bfseries 84} (2024) 833}
  [\href{https://arxiv.org/abs/2311.15280}{{\ttfamily 2311.15280}}].

\bibitem{Zhang:2024ebl}
S.~Zhang, K.-X. Chen and J.-C. Yang, \emph{{Detect anomalous quartic gauge
  couplings at muon colliders with quantum kernel k-means}},  9, 2024,
  \href{https://arxiv.org/abs/2409.07010}{{\ttfamily 2409.07010}}.

\bibitem{Zhu:2024own}
Y.~Zhu, W.~Zhuang, C.~Qian, Y.~Ma, D.~E. Liu, M.~Ruan et~al., \emph{{A Novel
  Quantum Realization of Jet Clustering in High-Energy Physics Experiments}},
  7, 2024, \href{https://arxiv.org/abs/2407.09056}{{\ttfamily 2407.09056}}.

\bibitem{Fadol:2022umw}
A.~Fadol, Q.~Sha, Y.~Fang, Z.~Li, S.~Qian, Y.~Xiao et~al., \emph{{Application
  of quantum machine learning in a Higgs physics study at the CEPC}},
  \href{https://doi.org/10.1142/S0217751X24500076}{\emph{Int. J. Mod. Phys. A}
  {\bfseries 39} (2024) 2450007}
  [\href{https://arxiv.org/abs/2209.12788}{{\ttfamily 2209.12788}}].

\bibitem{Wu:2020cye}
S.~L. Wu et~al., \emph{{Application of quantum machine learning using the
  quantum variational classifier method to high energy physics analysis at the
  LHC on IBM quantum computer simulator and hardware with 10 qubits}},
  \href{https://doi.org/10.1088/1361-6471/ac1391}{\emph{J. Phys. G} {\bfseries
  48} (2021) 125003} [\href{https://arxiv.org/abs/2012.11560}{{\ttfamily
  2012.11560}}].

\bibitem{Terashi:2020wfi}
K.~Terashi, M.~Kaneda, T.~Kishimoto, M.~Saito, R.~Sawada and J.~Tanaka,
  \emph{{Event Classification with Quantum Machine Learning in High-Energy
  Physics}}, \href{https://doi.org/10.1007/s41781-020-00047-7}{\emph{Comput.
  Softw. Big Sci.} {\bfseries 5} (2021) 2}
  [\href{https://arxiv.org/abs/2002.09935}{{\ttfamily 2002.09935}}].

\bibitem{Yang:2024bqw}
J.-C. Yang, S.~Zhang and C.-X. Yue, \emph{{A novel quantum machine learning
  classifier to search for new physics}},  10, 2024,
  \href{https://arxiv.org/abs/2410.18847}{{\ttfamily 2410.18847}}.

\bibitem{Feynman:1981tf}
R.~P. Feynman, \emph{{Simulating physics with computers}},
  \href{https://doi.org/10.1007/BF02650179}{\emph{Int. J. Theor. Phys.}
  {\bfseries 21} (1982) 467}.

\bibitem{Georgescu:2013oza}
I.~M. Georgescu, S.~Ashhab and F.~Nori, \emph{{Quantum Simulation}},
  \href{https://doi.org/10.1103/RevModPhys.86.153}{\emph{Rev. Mod. Phys.}
  {\bfseries 86} (2014) 153} [\href{https://arxiv.org/abs/1308.6253}{{\ttfamily
  1308.6253}}].

\bibitem{Bauer:2022hpo}
C.~W. Bauer et~al., \emph{{Quantum Simulation for High-Energy Physics}},
  \href{https://doi.org/10.1103/PRXQuantum.4.027001}{\emph{PRX Quantum}
  {\bfseries 4} (2023) 027001}
  [\href{https://arxiv.org/abs/2204.03381}{{\ttfamily 2204.03381}}].

\bibitem{Carena:2022kpg}
M.~Carena, H.~Lamm, Y.-Y. Li and W.~Liu, \emph{{Improved Hamiltonians for
  Quantum Simulations of Gauge Theories}},
  \href{https://doi.org/10.1103/PhysRevLett.129.051601}{\emph{Phys. Rev. Lett.}
  {\bfseries 129} (2022) 051601}
  [\href{https://arxiv.org/abs/2203.02823}{{\ttfamily 2203.02823}}].

\bibitem{Gustafson:2022xdt}
E.~J. Gustafson, H.~Lamm, F.~Lovelace and D.~Musk, \emph{{Primitive quantum
  gates for an SU(2) discrete subgroup: Binary tetrahedral}},
  \href{https://doi.org/10.1103/PhysRevD.106.114501}{\emph{Phys. Rev. D}
  {\bfseries 106} (2022) 114501}
  [\href{https://arxiv.org/abs/2208.12309}{{\ttfamily 2208.12309}}].

\bibitem{Lamm:2024jnl}
H.~Lamm, Y.-Y. Li, J.~Shu, Y.-L. Wang and B.~Xu, \emph{{Block encodings of
  discrete subgroups on a quantum computer}},
  \href{https://doi.org/10.1103/PhysRevD.110.054505}{\emph{Phys. Rev. D}
  {\bfseries 110} (2024) 054505}
  [\href{https://arxiv.org/abs/2405.12890}{{\ttfamily 2405.12890}}].

\bibitem{Carena:2024dzu}
M.~Carena, H.~Lamm, Y.-Y. Li and W.~Liu, \emph{{Quantum error thresholds for
  gauge-redundant digitizations of lattice field theories}},
  \href{https://doi.org/10.1103/PhysRevD.110.054516}{\emph{Phys. Rev. D}
  {\bfseries 110} (2024) 054516}
  [\href{https://arxiv.org/abs/2402.16780}{{\ttfamily 2402.16780}}].

\bibitem{Li:2023vwx}
Y.-Y. Li, M.~O. Sajid and J.~Unmuth-Yockey, \emph{{Lattice holography on a
  quantum computer}},
  \href{https://doi.org/10.1103/PhysRevD.110.034507}{\emph{Phys. Rev. D}
  {\bfseries 110} (2024) 034507}
  [\href{https://arxiv.org/abs/2312.10544}{{\ttfamily 2312.10544}}].

\bibitem{Cui:2019sfz}
X.~Cui, Y.~Shi and J.-C. Yang, \emph{{Circuit-based digital adiabatic quantum
  simulation and pseudoquantum simulation as new approaches to lattice gauge
  theory}}, \href{https://doi.org/10.1007/JHEP08(2020)160}{\emph{JHEP}
  {\bfseries 08} (2020) 160}
  [\href{https://arxiv.org/abs/1910.08020}{{\ttfamily 1910.08020}}].

\bibitem{Zou:2021pvl}
Y.-T. Zou, Y.-J. Bo and J.-C. Yang, \emph{{Optimize quantum simulation using a
  force-gradient integrator}},
  \href{https://doi.org/10.1209/0295-5075/135/10004}{\emph{EPL} {\bfseries 135}
  (2021) 10004} [\href{https://arxiv.org/abs/2103.05876}{{\ttfamily
  2103.05876}}].

\bibitem{Echevarria:2020wct}
M.~G. Echevarria, I.~L. Egusquiza, E.~Rico and G.~Schnell, \emph{{Quantum
  simulation of light-front parton correlators}},
  \href{https://doi.org/10.1103/PhysRevD.104.014512}{\emph{Phys. Rev. D}
  {\bfseries 104} (2021) 014512}
  [\href{https://arxiv.org/abs/2011.01275}{{\ttfamily 2011.01275}}].

\bibitem{Motta:2019yya}
M.~Motta, C.~Sun, A.~T.~K. Tan, M.~J.~O. Rourke, E.~Ye, A.~J. Minnich et~al.,
  \emph{{Determining eigenstates and thermal states on a quantum computer using
  quantum imaginary time evolution}},
  \href{https://doi.org/10.1038/s41567-019-0704-4}{\emph{Nature Phys.}
  {\bfseries 16} (2020) 205}
  [\href{https://arxiv.org/abs/1901.07653}{{\ttfamily 1901.07653}}].

\bibitem{Czajka:2021yll}
A.~M. Czajka, Z.-B. Kang, H.~Ma and F.~Zhao, \emph{{Quantum simulation of
  chiral phase transitions}},
  \href{https://doi.org/10.1007/JHEP08(2022)209}{\emph{JHEP} {\bfseries 08}
  (2022) 209} [\href{https://arxiv.org/abs/2112.03944}{{\ttfamily
  2112.03944}}].

\bibitem{Hasenfratz:1983ba}
P.~Hasenfratz and F.~Karsch, \emph{{Chemical Potential on the Lattice}},
  \href{https://doi.org/10.1016/0370-2693(83)91290-X}{\emph{Phys. Lett. B}
  {\bfseries 125} (1983) 308}.

\bibitem{deForcrand:2009zkb}
P.~de~Forcrand, \emph{{Simulating QCD at finite density}},
  \href{https://doi.org/10.22323/1.091.0010}{\emph{PoS} {\bfseries LAT2009}
  (2009) 010} [\href{https://arxiv.org/abs/1005.0539}{{\ttfamily 1005.0539}}].

\bibitem{Gattringer:2016kco}
C.~Gattringer and K.~Langfeld, \emph{{Approaches to the sign problem in lattice
  field theory}}, \href{https://doi.org/10.1142/S0217751X16430077}{\emph{Int.
  J. Mod. Phys. A} {\bfseries 31} (2016) 1643007}
  [\href{https://arxiv.org/abs/1603.09517}{{\ttfamily 1603.09517}}].

\bibitem{Fukushima:2013rx}
K.~Fukushima and C.~Sasaki, \emph{{The phase diagram of nuclear and quark
  matter at high baryon density}},
  \href{https://doi.org/10.1016/j.ppnp.2013.05.003}{\emph{Prog. Part. Nucl.
  Phys.} {\bfseries 72} (2013) 99}
  [\href{https://arxiv.org/abs/1301.6377}{{\ttfamily 1301.6377}}].

\bibitem{Busza:2018rrf}
W.~Busza, K.~Rajagopal and W.~van~der Schee, \emph{{Heavy Ion Collisions: The
  Big Picture, and the Big Questions}},
  \href{https://doi.org/10.1146/annurev-nucl-101917-020852}{\emph{Ann. Rev.
  Nucl. Part. Sci.} {\bfseries 68} (2018) 339}
  [\href{https://arxiv.org/abs/1802.04801}{{\ttfamily 1802.04801}}].

\bibitem{STAR:2010vob}
{\scshape STAR} collaboration, \emph{{An Experimental Exploration of the QCD
  Phase Diagram: The Search for the Critical Point and the Onset of
  De-confinement}},  7, 2010, \href{https://arxiv.org/abs/1007.2613}{{\ttfamily
  1007.2613}}.

\bibitem{Gupta:2011wh}
S.~Gupta, X.~Luo, B.~Mohanty, H.~G. Ritter and N.~Xu, \emph{{Scale for the
  Phase Diagram of Quantum Chromodynamics}},
  \href{https://doi.org/10.1126/science.1204621}{\emph{Science} {\bfseries 332}
  (2011) 1525} [\href{https://arxiv.org/abs/1105.3934}{{\ttfamily 1105.3934}}].

\bibitem{Andronic:2017pug}
A.~Andronic, P.~Braun-Munzinger, K.~Redlich and J.~Stachel, \emph{{Decoding the
  phase structure of QCD via particle production at high energy}},
  \href{https://doi.org/10.1038/s41586-018-0491-6}{\emph{Nature} {\bfseries
  561} (2018) 321} [\href{https://arxiv.org/abs/1710.09425}{{\ttfamily
  1710.09425}}].

\bibitem{Luo:2020pef}
X.~Luo, S.~Shi, N.~Xu and Y.~Zhang, \emph{{A Study of the Properties of the QCD
  Phase Diagram in High-Energy Nuclear Collisions}},
  \href{https://doi.org/10.3390/particles3020022}{\emph{Particles} {\bfseries
  3} (2020) 278} [\href{https://arxiv.org/abs/2004.00789}{{\ttfamily
  2004.00789}}].

\bibitem{STAR:2021iop}
{\scshape STAR} collaboration, \emph{{Cumulants and correlation functions of
  net-proton, proton, and antiproton multiplicity distributions in Au+Au
  collisions at energies available at the BNL Relativistic Heavy Ion
  Collider}}, \href{https://doi.org/10.1103/PhysRevC.104.024902}{\emph{Phys.
  Rev. C} {\bfseries 104} (2021) 024902}
  [\href{https://arxiv.org/abs/2101.12413}{{\ttfamily 2101.12413}}].

\bibitem{Stephans:2006tg}
G.~S.~F. Stephans, \emph{{critRHIC: The RHIC low energy program}},
  \href{https://doi.org/10.1088/0954-3899/32/12/S54}{\emph{J. Phys. G}
  {\bfseries 32} (2006) S447}
  [\href{https://arxiv.org/abs/nucl-ex/0607030}{{\ttfamily nucl-ex/0607030}}].

\bibitem{NA49-future:2006qne}
{\scshape NA49-future} collaboration, \emph{{Study of hadron production in
  hadron nucleus and nucleus nucleus collisions at the CERN SPS}},  11, 2006.

\bibitem{Gazdzicki:2008kk}
{\scshape NA61/SHINE} collaboration, \emph{{Ion Program of Na61/Shine at the
  CERN SPS}}, \href{https://doi.org/10.1088/0954-3899/36/6/064039}{\emph{J.
  Phys. G} {\bfseries 36} (2009) 064039}
  [\href{https://arxiv.org/abs/0812.4415}{{\ttfamily 0812.4415}}].

\bibitem{Luo:2017faz}
X.~Luo and N.~Xu, \emph{{Search for the QCD Critical Point with Fluctuations of
  Conserved Quantities in Relativistic Heavy-Ion Collisions at RHIC : An
  Overview}}, \href{https://doi.org/10.1007/s41365-017-0257-0}{\emph{Nucl. Sci.
  Tech.} {\bfseries 28} (2017) 112}
  [\href{https://arxiv.org/abs/1701.02105}{{\ttfamily 1701.02105}}].

\bibitem{Stephanov:2004wx}
M.~A. Stephanov, \emph{{QCD Phase Diagram and the Critical Point}},
  \href{https://doi.org/10.1143/PTPS.153.139}{\emph{Prog. Theor. Phys. Suppl.}
  {\bfseries 153} (2004) 139}
  [\href{https://arxiv.org/abs/hep-ph/0402115}{{\ttfamily hep-ph/0402115}}].

\bibitem{Fodor:2004nz}
Z.~Fodor and S.~D. Katz, \emph{{Critical point of QCD at finite T and mu,
  lattice results for physical quark masses}},
  \href{https://doi.org/10.1088/1126-6708/2004/04/050}{\emph{JHEP} {\bfseries
  04} (2004) 050} [\href{https://arxiv.org/abs/hep-lat/0402006}{{\ttfamily
  hep-lat/0402006}}].

\bibitem{Gavai:2004sd}
R.~V. Gavai and S.~Gupta, \emph{{The Critical end point of QCD}},
  \href{https://doi.org/10.1103/PhysRevD.71.114014}{\emph{Phys. Rev. D}
  {\bfseries 71} (2005) 114014}
  [\href{https://arxiv.org/abs/hep-lat/0412035}{{\ttfamily hep-lat/0412035}}].

\bibitem{Yang:2022zob}
J.-C. Yang, X.-T. Chang and J.-X. Chen, \emph{{Study of the Roberge-Weiss phase
  caused by external uniform classical electric field using lattice QCD
  approach}}, \href{https://doi.org/10.1007/JHEP10(2022)053}{\emph{JHEP}
  {\bfseries 10} (2022) 053}
  [\href{https://arxiv.org/abs/2207.11796}{{\ttfamily 2207.11796}}].

\bibitem{Yang:2023vsw}
J.-C. Yang and X.-G. Huang, \emph{{QCD on Rotating Lattice with Staggered
  Fermions}},  7, 2023.

\bibitem{Yang:2023zzx}
J.-C. Yang, X.~Zhang and J.-X. Chen, \emph{{Study of the effects of external
  imaginary electric field and chiral chemical potential on quark matter}},
  \href{https://doi.org/10.1140/epjc/s10052-024-13069-x}{\emph{Eur. Phys. J. C}
  {\bfseries 84} (2024) 746}
  [\href{https://arxiv.org/abs/2309.09281}{{\ttfamily 2309.09281}}].

\bibitem{Thirring:1958in}
W.~E. Thirring, \emph{{A Soluble relativistic field theory?}},
  \href{https://doi.org/10.1016/0003-4916(58)90015-0}{\emph{Annals Phys.}
  {\bfseries 3} (1958) 91}.

\bibitem{Sachs:1995dm}
I.~Sachs and A.~Wipf, \emph{{Generalized Thirring models}},
  \href{https://doi.org/10.1006/aphy.1996.0077}{\emph{Annals Phys.} {\bfseries
  249} (1996) 380} [\href{https://arxiv.org/abs/hep-th/9508142}{{\ttfamily
  hep-th/9508142}}].

\bibitem{Mishra:2019xbh}
C.~Mishra, S.~Thompson, R.~Pooser and G.~Siopsis, \emph{{Quantum computation of
  an interacting fermionic model}},
  \href{https://doi.org/10.1088/2058-9565/ab8f63}{\emph{Quantum Sci. Technol.}
  {\bfseries 5} (2020) 035010}
  [\href{https://arxiv.org/abs/1912.07767}{{\ttfamily 1912.07767}}].

\bibitem{Hagen1967}
C.~R. Hagen, \emph{{New solutions of the thirring model}},
  \href{https://doi.org/10.1007/BF02712329}{\emph{Nuovo Cimento B (1965-1970)}
  {\bfseries 51} (1967) 169}.

\bibitem{Korepin_Bogoliubov_Izergin_1993}
V.~E. Korepin, N.~M. Bogoliubov and A.~G. Izergin, \emph{Quantum Inverse
  Scattering Method and Correlation Functions}, Cambridge Monographs on
  Mathematical Physics. Cambridge University Press, 1993,
  \href{https://doi.org/10.1017/CBO9780511628832}{10.1017/CBO9780511628832}.

\bibitem{Alvarez-Estrada:1997vmv}
R.~F. Alvarez-Estrada and A.~Gomez~Nicola, \emph{{The Schwinger and Thirring
  models at finite chemical potential and temperature}},
  \href{https://doi.org/10.1103/PhysRevD.57.3618}{\emph{Phys. Rev. D}
  {\bfseries 57} (1998) 3618}
  [\href{https://arxiv.org/abs/hep-th/9710227}{{\ttfamily hep-th/9710227}}].

\bibitem{PintoBarros:2018bzz}
J.~C. Pinto~Barros, M.~Dalmonte and A.~Trombettoni, \emph{{String tension and
  robustness of confinement properties in the Schwinger-Thirring model}},
  \href{https://doi.org/10.1103/PhysRevD.100.036009}{\emph{Phys. Rev. D}
  {\bfseries 100} (2019) 036009}
  [\href{https://arxiv.org/abs/1808.00444}{{\ttfamily 1808.00444}}].

\bibitem{Coleman:1974bu}
S.~R. Coleman, \emph{{The Quantum Sine-Gordon Equation as the Massive Thirring
  Model}}, \href{https://doi.org/10.1103/PhysRevD.11.2088}{\emph{Phys. Rev. D}
  {\bfseries 11} (1975) 2088}.

\bibitem{Banuls:2017evv}
M.~C. Ba\~nuls, K.~Cichy, Y.-J. Kao, C.~J.~D. Lin, Y.-P. Lin and D.~T.-L. Tan,
  \emph{{Tensor Network study of the (1+1)-dimensional Thirring Model}},
  \href{https://doi.org/10.1051/epjconf/201817511017}{\emph{EPJ Web Conf.}
  {\bfseries 175} (2018) 11017}
  [\href{https://arxiv.org/abs/1710.09993}{{\ttfamily 1710.09993}}].

\bibitem{Gross:1974jv}
D.~J. Gross and A.~Neveu, \emph{{Dynamical Symmetry Breaking in Asymptotically
  Free Field Theories}},
  \href{https://doi.org/10.1103/PhysRevD.10.3235}{\emph{Phys. Rev. D}
  {\bfseries 10} (1974) 3235}.

\bibitem{Chen:2010gda}
X.~Chen, Z.~C. Gu and X.~G. Wen, \emph{{Local unitary transformation,
  long-range quantum entanglement, wave function renormalization, and
  topological order}},
  \href{https://doi.org/10.1103/PhysRevB.82.155138}{\emph{Phys. Rev. B}
  {\bfseries 82} (2010) 155138}
  [\href{https://arxiv.org/abs/1004.3835}{{\ttfamily 1004.3835}}].

\bibitem{Kogut:1974ag}
J.~B. Kogut and L.~Susskind, \emph{{Hamiltonian Formulation of Wilson's Lattice
  Gauge Theories}}, \href{https://doi.org/10.1103/PhysRevD.11.395}{\emph{Phys.
  Rev. D} {\bfseries 11} (1975) 395}.

\bibitem{Kluberg-Stern:1983lmr}
H.~Kluberg-Stern, A.~Morel, O.~Napoly and B.~Petersson, \emph{{Flavors of
  Lagrangian Susskind Fermions}},
  \href{https://doi.org/10.1016/0550-3213(83)90501-1}{\emph{Nucl. Phys. B}
  {\bfseries 220} (1983) 447}.

\bibitem{Morel:1984di}
A.~Morel and J.~P. Rodrigues, \emph{{How to Extract {QCD} Baryons From a
  Lattice Theory With Staggered Fermions}},
  \href{https://doi.org/10.1016/0550-3213(84)90371-7}{\emph{Nucl. Phys. B}
  {\bfseries 247} (1984) 44}.

\bibitem{JW}
P.~{Jordan} and E.~{Wigner}, \emph{{{\"U}ber das Paulische
  {\"A}quivalenzverbot}},
  \href{https://doi.org/10.1007/BF01331938}{\emph{Zeitschrift fur Physik}
  {\bfseries 47} (1928) 631}.

\bibitem{Delepine:1997bz}
D.~Delepine, R.~Gonzalez~Felipe and J.~Weyers, \emph{{Equivalence of the
  sine-Gordon and massive Thirring models at finite temperature}},
  \href{https://doi.org/10.1016/S0370-2693(97)01436-6}{\emph{Phys. Lett. B}
  {\bfseries 419} (1998) 296}
  [\href{https://arxiv.org/abs/hep-th/9709039}{{\ttfamily hep-th/9709039}}].

\bibitem{Shen:2024fcj}
Y.~Shen, K.~Klymko, E.~Rabani, N.~M. Tubman, D.~Camps, R.~Van~Beeumen et~al.,
  \emph{{Simple Diagonal Designs with Reconfigurable Real-Time Circuits}},  1,
  2024.

\bibitem{Lloyd:1996aai}
S.~Lloyd, \emph{{Universal Quantum Simulators}},
  \href{https://doi.org/10.1126/science.273.5278.1073}{\emph{Science}
  {\bfseries 273} (1996) 1073}.

\bibitem{qiskit2024}
A.~Javadi-Abhari, M.~Treinish, K.~Krsulich, C.~J. Wood, J.~Lishman, J.~Gacon
  et~al., \emph{Quantum computing with {Q}iskit},  2024.
\newblock 10.48550/arXiv.2405.08810.

\bibitem{Kolotouros:2024inn}
I.~Kolotouros, D.~Joseph and A.~K. Narayanan, \emph{{Accelerating quantum
  imaginary-time evolution with random measurements}},
  \href{https://doi.org/10.1103/PhysRevA.111.012424}{\emph{Phys. Rev. A}
  {\bfseries 111} (2025) 012424}
  [\href{https://arxiv.org/abs/2407.03123}{{\ttfamily 2407.03123}}].

\bibitem{Nishi:2021xhf}
H.~Nishi, T.~Kosugi and Y.-i. Matsushita, \emph{{Implementation of quantum
  imaginary-time evolution method on NISQ devices by introducing nonlocal
  approximation}}, \href{https://doi.org/10.1038/s41534-021-00409-y}{\emph{npj
  Quantum Inf.} {\bfseries 7} (2021) 85}.

\bibitem{Kumar:2024kij}
S.~Kumar and C.~M. Wilmott, \emph{{Generalising quantum imaginary time
  evolution to solve linear partial differential equations}},
  \href{https://doi.org/10.1038/s41598-024-70423-5}{\emph{Sci. Rep.} {\bfseries
  14} (2024) 20156} [\href{https://arxiv.org/abs/2405.01313}{{\ttfamily
  2405.01313}}].

\bibitem{Gomes:2021ckn}
N.~Gomes, A.~Mukherjee, F.~Zhang, T.~Iadecola, C.-Z. Wang, K.-M. Ho et~al.,
  \emph{{Adaptive Variational Quantum Imaginary Time Evolution Approach for
  Ground State Preparation}},
  \href{https://doi.org/10.1002/qute.202100114}{\emph{Adv. Quantum Technol.}
  {\bfseries 4} (2021) 2100114}
  [\href{https://arxiv.org/abs/2102.01544}{{\ttfamily 2102.01544}}].

\bibitem{Silva:2021jaj}
T.~d.~L. Silva, M.~M. Taddei, S.~Carrazza and L.~Aolita, \emph{{Fragmented
  imaginary-time evolution for early-stage quantum signal processors}},
  \href{https://doi.org/10.1038/s41598-023-45540-2}{\emph{Sci. Rep.} {\bfseries
  13} (2023) 18258} [\href{https://arxiv.org/abs/2110.13180}{{\ttfamily
  2110.13180}}].

\bibitem{Hejazi:2023fiq}
K.~Hejazi, M.~Motta and G.~K.-L. Chan, \emph{{Adiabatic quantum imaginary time
  evolution}},
  \href{https://doi.org/10.1103/PhysRevResearch.6.033084}{\emph{Phys. Rev.
  Res.} {\bfseries 6} (2024) 033084}
  [\href{https://arxiv.org/abs/2308.03292}{{\ttfamily 2308.03292}}].

\bibitem{Gluza:2024lqq}
M.~Gluza, J.~Son, B.~H. Tiang, Y.~Suzuki, Z.~Holmes and N.~H.~Y. Ng,
  \emph{{Double-bracket quantum algorithms for quantum imaginary-time
  evolution}},  12, 2024.

\end{thebibliography}\endgroup
\bibliographystyle{JHEP}

\end{document}